\documentclass{format170x240multiauthor-arXiv}
\usepackage[T1]{fontenc}
\usepackage{makeidx}
   \makeindex
\usepackage{amsmath}
\usepackage{textcomp}
\usepackage{graphicx}
\usepackage{cite} 

\usepackage{amssymb}	

\newcommand{\ket}[1]{|#1 \rangle}		
\newcommand{\um}{\textmu m }			
\newcommand{\uK}{\textmu K }			
\newcommand{\pf}{\rm{^{40}K}}   		
\newcommand{\re}{\rm{^{87}Rb}}	        	

\begin{document}

\frontmatter
\tableofcontents
\mainmatter

\chapter{Fermions on atom chips}
\label{YYZ:chapter:fermions}
\chapterauthor{Marcius H.T. Extavour, Lindsay J. LeBlanc, Jason
McKeever, Alma B. Bardon, \mbox{Seth Aubin}, Stefan Myrskog, 
Thorsten Schumm and Joseph H. Thywissen}

\section{Introduction}
Degenerate Fermi gases (DFGs) earned their place at the leading edge
of degenerate quantum gas research with the first demonstration of
an atomic $\pf$ DFG in 1999 \cite{YYZ:DeMarco:firstDFG}.  Since then,
DFGs have been central to many important advances in the field,
including boson-fermion mixtures, strongly interacting fermion-fermion
spin mixtures, Bose-Einstein condensates (BECs) of fermion
dimers, Bardeen-Cooper-Shrieffer (BCS) -type superfluidity, BEC-BCS
crossover physics, Bose-Fermi heteronuclear molecules, and fermions in
optical lattices (see \cite{YYZ:manyBody-RMP:2008} and references
therein).\index{Fermi gas} \index{DFG} \index{$\pf$} \index{potassium}
\index{BEC-BCS crossover} \index{degenerate Fermi gas}

Not long after the first DFG was produced, efficient loading of cold
atoms into atom chip microtraps enabled the first demonstration of
Bose-Einstein condensation on an atom chip \cite{YYZ:Ott:chipBEC,
YYZ:Hansel:chipBEC}. In subsequent years, research efforts in DFGs and
atom chips progressed independently, but were combined with the 2006
demonstration in Toronto of a DFG of $\pf$ on an atom chip
\cite{YYZ:Aubin:chip-DFG}.

There are several attractive technical advantages in using atom chips
for fermions. Large collision rates, made possible by the strong
confinement of microtraps relative to conventional, macroscopic
traps\footnote{We distinguish between ``macroscopic'' traps generated
by centimetre-scale external magnetic field coils, and chip-based
microtraps.}, permit rapid sympathetic evaporative cooling to quantum
degeneracy.  This obviates the need for minute-scale vacuum lifetimes,
multi-chamber vacuum systems and Zeeman slowers. Dramatically shorter
experimental cycle times from atomic vapour to DFG also become
possible, a point of practical value in day-to-day laboratory
research.

Atom chips also offer a wide array of techniques for trapping and
manipulating ultra-cold atoms using a single micro-fabricated device,
including magnetostatic, electrostatic, and dynamic radio-frequency
(RF) and microwave dressed potentials, and integrated optical
potentials. With these tools, discussed extensively in this book, atom
chips are well poised to advance active research in fermionic
many-body systems:

\begin{description} 
\item[Fermions in one dimension] Atom chips excel at producing
anisotropic, high aspect ratio magnetic potentials.  They are
particularly well-suited to the study of one-dimensional fermion
physics such as Luttinger liquids \cite{YYZ:Mathey:Luttinger,
YYZ:Rauf:Luttinger-nanotubes, YYZ:Lieb:1D}, confinement-induced
molecule formation \cite{YYZ:Moritz:1D-fermion-molecs} and spin-charge
separation \cite{YYZ:Recati:spin-charge}. 
\index{spin-charge separation} \index{Luttinger liquid}

\item[Interference and coherence] RF-dressed double-well potentials on
atom chips \cite{YYZ:Schumm:doubleBEC,YYZ:Hofferberth-RF1,
YYZ:Lesanovsky:RF,YYZ:GBJo:dw1,YYZ:Extavour:dual-degen} have only
recently been applied to ultra-cold fermions
\cite{YYZ:Extavour:dual-degen}. This combination is an exciting step
in the direction of fermion quantum atom optics, including
interferometry \cite{YYZ:Yurke:fermi-interf}, mesoscopic quantum
pumping circuit simulations \cite{YYZ:Das:q-pump}, antibunching
\cite{YYZ:Jeltes:bunch-antibunch}, and number statistics
\cite{YYZ:Tran:fermi-n-flucts1, YYZ:Tran:fermi-n-flucts2,
YYZ:Budde:fermi-num-distn}.\index{RF-dressed potentials}

\item[Strongly interacting Fermi gases] Spin-independent optical
potentials can trap the spin mixtures necessary for strong
inter-particle interactions in fermions. Optical traps near the
surface of an atom chip \cite{YYZ:Wang:chip-optical-interf,
YYZ:Colombe:chip-cavity} provide opportunities to combine
strongly interacting DFGs with near-field RF and microwave
probes.
\end{description}

In this chapter we review our recent and ongoing work with Fermi gases
and Bose-Fermi mixtures on an atom chip. The chapter begins with a
review of statistical and thermodynamic properties of the ideal,
non-interacting Fermi gas.  After a brief description of our atom chip
and its capabilities, we discuss our experimental approach to
producing a $\pf$ DFG and a \mbox{$\pf$--$\re$} DFG-BEC mixture.  In
doing so, we describe the factors affecting the loading efficiency of
the atom chip microtrap, and review our rapid sympathetic evaporation
to degeneracy. This is followed by a discussion of species selectivity
in RF manipulation of the \mbox{$\pf$--$\re$} mixture, which we
explore in the context of sympathetic evaporative cooling and
RF-dressed adiabatic double-well potentials.  Next, we describe the
incorporation of a crossed-beam dipole trap into the atom chip setup,
with which we generate and manipulate strongly interacting spin
mixtures of $\pf$.  Finally, we conclude with a brief discussion of
future research directions with DFGs and atom chips.
\index{sympathetic evaporative cooling} \index{RF-dressed potentials}

\section{Theory of ideal Fermi gases}
\label{YYZ:sec:DFG-background}
Ultra-cold Fermi gases differ from ultra-cold Bose gases in their
simplest theoretical description in two important ways: first, there
is no macroscopic occupation of the single-particle ground state;
second, spin-polarized Fermi gases are completely non-interacting at
ultra-cold temperatures \cite{YYZ:CastinFermiBox, YYZ:BruunClark}.
Ideal thermodynamic functions are thus excellent descriptors of cold
spin-polarized fermions, even as $T \to 0$.  In this section we review
fermion thermodynamics, calculate trapped density distributions, and
discuss observable signatures of Fermi degeneracy. 
\index{Fermi degeneracy}

\subsection{Thermodynamics}
\label{YYZ:sec:DFG-thermo}
\begin{figure} 
\begin{center}
\includegraphics[width=8cm]{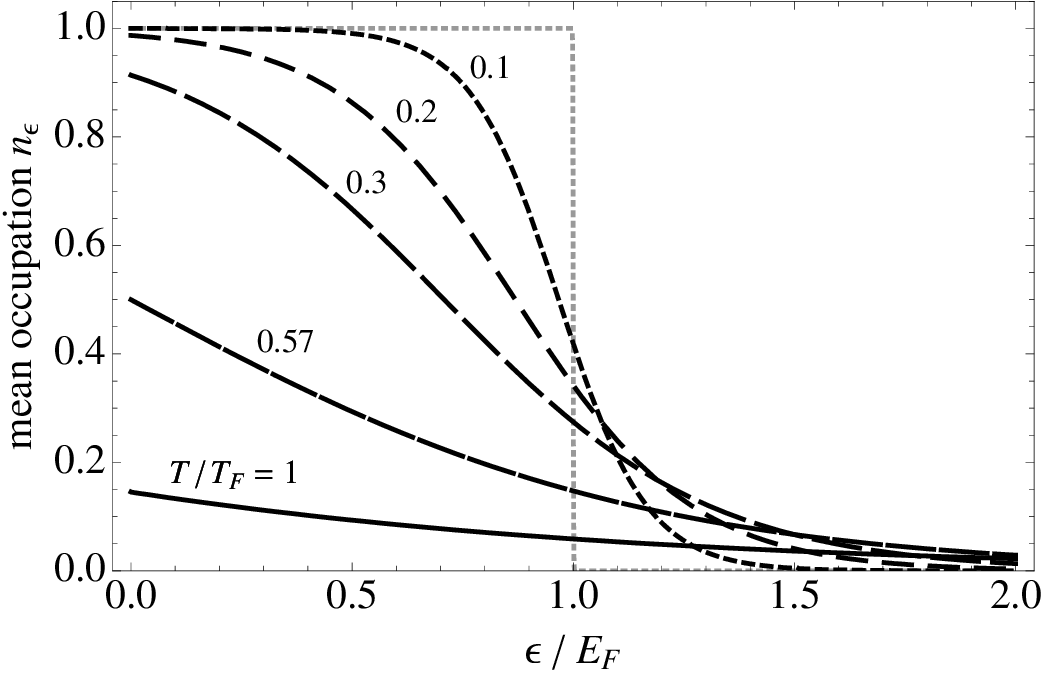} 
\Caption{Universal curves of the mean Fermi-Dirac 
occupation of single particle states $\epsilon$ vs. 
$\epsilon/E_F$, shown at select values of $T/T_F$. The grey, dotted
line indicates the $T=0$ filled Fermi sea (see text).}
\label{YYZ:fig:FD-occupation} 
\end{center} 
\end{figure}

In the grand canonical ensemble description of an ideal Fermi gas, the
mean occupation number of the single-particle energy state $\epsilon$
is 
\begin{equation} \label{YYZ:eq:occn-number} 
n_\epsilon =
\frac{1}{e^{\beta(\epsilon - \mu)} +1} = 
\frac{1}{\mathcal{Z}^{-1} e^{\beta \epsilon} +1},
\end{equation} 
where $\beta \equiv 1/k_B T$, $k_B$ is the Boltzmann constant, $\mu$
is the chemical potential of the gas, and $\mathcal{Z} \equiv e^{\beta
\mu}$ is the fugacity. The mean occupation number is bounded $0 \leq
n_\epsilon \leq 1$, which is a statement of the Pauli exclusion
principle.  Figure~\ref{YYZ:fig:FD-occupation} shows the occupation
number as a function of single-particle energy for various
temperatures.  The $T=0$ ideal Fermi gas is characterized by a filled
``Fermi sea''\index{Fermi sea}: all energy levels $\epsilon$ for which
$\epsilon \leq E_F$ are fully occupied ($n_\epsilon = 1$), while those
for which $\epsilon > E_F$ are empty ($n_\epsilon = 0$).  The Fermi
energy $E_F$ is equal to the chemical potential $\mu$ at $T=0$. At
high temperatures the gas is described by a Boltzmann-like
distribution \cite{YYZ:Pathria:SM}.  \index{Fermi-Dirac distribution}
\index{canonical ensemble} \index{fugacity}

Since a trapped gas is not in contact with number or energy reservoirs
in experiments, we ignore fluctuations in the total number and energy
predicted by the grand canonical ensemble description, taking $N$ and
$E$ to be the \emph{average} total number and total energy,
respectively. These can be calculated using the discrete sums
\begin{equation} \label{YYZ:eq:discretesum} N  =  \sum_\epsilon
n_\epsilon \hspace{0.5in} \mbox{and} \hspace{0.5in} E  = \sum_\epsilon
\epsilon \, n_\epsilon, \end{equation}
where the sums run over all discrete states.  In the limit of a large
number of occupied states we can take the continuum limit,
writing 
\begin{equation} \label{YYZ:eq:continuumlimit} 
N  = \int_{\epsilon=0}^{\infty}  g(\epsilon) n_\epsilon \, d\epsilon
\hspace{0.5in} \mbox{and} \hspace{0.5in} 
E = \int_{\epsilon=0}^{\infty}  g(\epsilon)  \epsilon \, n_\epsilon \,
d\epsilon, 
\end{equation}
where $g(\epsilon) = \epsilon^2/2(\hbar \bar{\omega})^3$ is the energy
density of states \index{density of states} for a harmonically trapped
gas in three-dimensions, and $\bar{\omega} \equiv (\omega_x \omega_y
\omega_z)^{1/3}$ is the geometric mean harmonic trap frequency.

Integrals of this type are of the form
\begin{equation} \label{YYZ:eq:FermiIntegral}
\int_0^\infty a\,^{n-1} da \frac{1}{c^{-1} e \, ^a + 1} = 
\Gamma(n) f_{n}(C),
\end{equation}		
where $f_n(C) =  -\text{Li}_n(-C)$, and $\text{Li}_{n}(C) =
\sum_{j=1}^{\infty} C^j / j^n$ is a polylogarithmic function.
\index{polylogarithm}  For $n=1,$ $f_1=\ln{(1+C)}$.  Using
eq.~\ref{YYZ:eq:FermiIntegral}, we find that the total number and
energy are
\begin{equation} \label{YYZ:eq:NandE}
N = (\beta \hbar \bar{\omega})^{-3} f_3(\mathcal{Z})  
\hspace{0.5in} \mbox{and} \hspace{0.5in} 
E = 3 k_B T (\beta \hbar \bar{\omega})^{-3}   f_4(\mathcal{Z}).
\end{equation}

The number and energy at zero temperature can be found using the zero
temperature limit of the Fermi function\index{Fermi function},
\begin{equation} \label{YYZ:eq:zeroTlimit}	
\lim_{T \rightarrow 0} f_n(\mathcal{Z}) = 
\frac{(\beta \mu)^n}{\Gamma (n+1)}.
\end{equation}

As mentioned above, the Fermi energy $E_F$ is defined as the zero
temperature limit of $\mu$. For convenience, we will also refer to the
``Fermi temperature'' \index{Fermi temperature} $T_F \equiv E_F/k_B$,
even though this temperature does not correspond to a phase
transition, as is the case for $T_c$ of Bose gases. Re-writing
eq.~\ref{YYZ:eq:NandE} in terms of $E_F$, we find 
\begin{equation} \label{YYZ:eq:NandEzeroT} 
N =  \frac{1}{6} \left( \frac{E_F}{\hbar \bar{\omega}} \right)^3 
\hspace{0.5in} \mbox{and} \hspace{0.5in} 
E = \frac{3}{5} N \, E_F.  
\end{equation}

The chemical potential and fugacity \index{fugacity} at finite
temperature can be found numerically by solving
\begin{equation} 6 f_3(\mathcal{Z}) = (\beta E_F)^3. \end{equation}
Using the Sommerfeld expansion \index{Sommerfeld expansion} of the
polylogarithms, one obtains low- and high-temperature approximations
to the chemical potential in a three-dimensional harmonic trap
\cite{YYZ:ButtsRokhsar}:
\begin{equation} \label{YYZ:eq:mu-approx}
\mu \approx \left\{ \begin{array}{ll}
E_F \bigg[ 1 - \frac{\pi^2}{3} \left( \frac{k_B T}{E_F} \right)^2 \bigg]  
& \mbox{for $ k_B T \ll E_F $, \; and} \\
- k_B T \ln \bigg[ 6 \left( \frac{k_B T}{E_F} \right)^3 \bigg]
  & \mbox{for $ k_B T \gg E_F$.} \end{array} \right.
\end{equation}

\paragraph{Low dimensionality}
Under certain conditions, a $T=0$ Fermi gas in an anisotropic magnetic
trap having $\omega_\perp \gg \omega_\parallel$ may become effectively
one-dimensional. If the atom number and temperature are such that $E_F
< \hbar \omega_\perp$, the transverse degrees of freedom are ``frozen
out'' and fermions occupy only the longitudinal energy levels of the
trap.  The number of fermions that can occupy the 1D states at $T=0$
is equal to the aspect ratio of the trap: $N_{1D} = \omega_\perp /
\omega_\parallel$.  This scenario is especially relevant to atom chip
micromagnetic traps, whose aspect ratios can be on the order of 10$^2$
to 10$^4$ \cite{YYZ:Reichel:chip-TonksGas}.
\index{fermions in one dimension}

\subsection{Density distribution} \label{YYZ:sec:dens-distn}
Apart from the choice of $g(\epsilon)$, many of the expressions
derived in sec.~\ref{YYZ:sec:DFG-thermo} resemble the textbook
treatment of a uniform Fermi gas. In this section, we calculate the
non-uniform position and momentum distributions of \emph{trapped}
fermions.  The position distribution is observable \emph{in situ} (with
sufficient spatial resolution), while the momentum distribution is
observable in time-of-flight. We calculate these distributions by two
different conceptual starting points: first, using semi-classical
integrals; and second, using the local density
approximation.\index{local density approximation} 

\paragraph{Semi-classical approximation} We can express energy as a
semi-classical function of position and momentum, $\epsilon = p^2 / 2M
+ U(\mathbf{r})$, where $U(\mathbf{r})$ is the trapping potential and
$M$ is the atomic mass. Using the distribution of
eq.~\ref{YYZ:eq:occn-number}, we integrate over momentum degrees of
freedom to find the position distribution: \index{semi-classical
approximation}
\begin{equation} \label{YYZ:eq:positionintegral}
n(\mathbf{r}) = \int \frac{d^3 p}{(2 \pi \hbar)^3} 
[\mathcal{Z}^{-1} e^{\beta \epsilon(\mathbf{r}, \mathbf{p})} + 1]^{-1},
\end{equation}
in which we have used the semi-classical phase-space volume of one
quantum state, $(2 \pi \hbar)^3$.  Integration using
eq.~\ref{YYZ:eq:FermiIntegral} yields
\begin{equation} \label{YYZ:eq:positionFiniteT}
n(\mathbf{r}) = \Lambda_T^{-3} f_{3/2} 
(\mathcal{Z}e^{-\beta U(\mathbf{r})}),
\end{equation}
where $\Lambda_T =  \sqrt{2\pi \hbar^2 \beta / M}$ is the thermal de
Broglie wavelength \index{thermal de Broglie wavelength}. Unlike the
corresponding expression for ideal bosons,
eq.~\ref{YYZ:eq:positionFiniteT} is valid \emph{at all temperatures}.
The difference lies in the fact that we have ignored the occupation of
the single-particle ground state in taking the continuum limit (see
eq.~\ref{YYZ:eq:continuumlimit}), evidenced by the vanishing  density
of states $g(\epsilon)$ for $\epsilon=0$.  This does not pose a
problem for fermions, for which the occupation of the ground state
$\mathcal{Z}/(1+\mathcal{Z}) \leq 1$. For bosons, however, this limit
completely ignores the condensed fraction, whose contribution to the
thermodynamics must be reintroduced manually \cite{YYZ:Pathria:SM}.
\index{semi-classical approximation}

\paragraph{Local density approximation} 
\index{local density approximation} 
An alternate conceptual approach to the calculation of inhomogeneous
distributions is the local density approximation. We start with
expressions for a \emph{uniform} Fermi gas, and assume that local
properties can be described by a local chemical potential $\mu -
U(\mathbf{r})$ and local fugacity $\mathcal{Z}e^{-\beta
U(\mathbf{r})}$. This implies that long-range properties of the Fermi
gas may be ignored, unlike in a BEC, which exhibits long-range phase
coherence.  Since the density of a uniform degenerate Fermi gas is
\begin{equation} \label{YYZ:eq:densityUniform} 
n_\mathrm{uniform} =
\Lambda_T^{-3} f_{3/2} (\mathcal{Z})  \stackrel{T=0}{\longrightarrow} 
\frac{1}{6 \pi^2} \left[ \frac{2 M}{\hbar^2} E_F \right]^{3/2},
\end{equation}
we immediately recover eq.~\ref{YYZ:eq:positionFiniteT} and its
zero-temperature limit. In fact, the local density approximation and
the semi-classical approach generally yield identical results for
non-interacting fermions \cite{YYZ:CastinFermiBox}.
\index{local density approximation}\index{degenerate Fermi gas}

Specializing to the case of a three-dimensional harmonic oscillator
potential $U(\mathbf{r}) = \tfrac{1}{2} M 
(\omega_x^2 x^2 + \omega_y^2 y^2 + \omega_z^2 z^2)$ we obtain
\begin{equation} \label{YYZ:eq:positionSHO} 
n(\mathbf{r}) = \Lambda_T^{-3} f_{3/2} \bigg(
\mathcal{Z}\exp{\big[-\frac{\beta M}{2} 
(\omega_x^2 x^2 + \omega_y^2 y^2 + \omega_z^2 z^2)\big]}\bigg).
\end{equation}
At zero temperature, 
\begin{equation} \label{YYZ:eq:TFpositionZeroT}
n(\mathbf{r}) = \frac{8 N}{\pi^2 {\bar{R}}^3_{TF}} 
\left[ 1 - \frac{x^2}{X_{TF}^2}- \frac{y^2}{Y_{TF}^2} - 
\frac{z^2}{Z_{TF}^2} \right]^{3/2} 
\Theta \left(1 - \frac{x^2}{X_{TF}^2}- 
\frac{y^2}{Y_{TF}^2} - \frac{z^2}{Z_{TF}^2} \right)
\end{equation}
where $\bar{R}_{TF} = \sqrt{2 E_F / M \bar{\omega}^2 }$ is the mean
Thomas-Fermi radius \index{Thomas-Fermi radius} of the cloud, $X_{TF}
= \sqrt{2 E_F / M \omega_x^2 }$ \emph{etc.} are the Thomas-Fermi
lengths along each trapping axis, and $\Theta(\cdots)$ is the
Heaviside step function. 

The momentum distribution can also be calculated using either the
local density or semi-classical approach. Experimentally, we 
observe the momentum distribution in a time-of-flight density image,
for which the distribution is obtained by rescaling all frequencies
$\omega_i \to \omega_i \sqrt{1 + \omega_i^2 t^2}$ in
eq.~\ref{YYZ:eq:positionSHO} along the direction $i \in \{x,y,z\}$,
and renormalizing to conserve particle number \cite{YYZ:BruunClark}.
\index{local density approximation} 

\subsection{Crossover to Fermi degeneracy}\index{Fermi degeneracy}
\label{YYZ:sec:fermi-degen}
The $T=0$ filled Fermi sea \index{Fermi sea} is quantum degenerate
\index{quantum degeneracy} in the sense that it represents the
absolute many-particle ground state of this non-interacting system.
The meaning of the term ``degenerate'' here should not be confused
with its more conventional meaning for a gas of bosons, for which
degeneracy implies multiple or macroscopic occupation of the
single-particle ground state. Multiple occupancy is forbidden for
fermions, and so they are always ``non-degenerate'' in the more
conventional sense of the word.

What, then, is the nature of the transition to quantum degeneracy in
fermions? In contrast to the boson case, there is no phase
transition into or out of the filled Fermi sea\index{Fermi sea}. As is
the case with bosons, however, high-temperature expansions for
thermodynamic quantities fail around $\mathcal{Z}=1$.  At lower
temperatures the behaviour differs dramatically from the predictions
of classical, Boltzmann statistics. Whereas for low-temperature ideal
bosons $\mathcal{Z} \rightarrow 0^-$ as $T \rightarrow 0$, for fermions
$\mathcal{Z}\rightarrow \infty$ with the scaling $\mathcal{Z}\approx
e^{\beta E_F}$, as implied by eq.~\ref{YYZ:eq:mu-approx}. It is
interesting to note the quantitative relationship between fugacity and
degeneracy for fermions:
\begin{equation} \label{YYZ:eq:degen-param} 
n_0 \Lambda_T^3 = f_{3/2}(\mathcal{Z}),
\end{equation}
where $n_0 \equiv n(0)$ is the central density of the cloud.
Thus  $n_0 \Lambda_T^3 \simeq 0.77$ when $\mathcal{Z}=1$ for
fermions, which occurs at $T \simeq 0.57 T_F$.  By comparison, $
n_0 \Lambda_T^3 \simeq 2.61$ when $\mathcal{Z}=1$ for bosons, at
$T=T_c$.

The lack of a marked phase transition raises the question of what an
experimental signature of Fermi degeneracy might be. Unlike a BEC,
the non-interacting DFG has an isotropic momentum distribution in
time-of-flight, even when released from an anisotropic trap
\cite{YYZ:ButtsRokhsar}.  Thus the aspect ratio of the cold cloud
cannot be a signature of degeneracy. Instead, observations of Fermi
degeneracy rely on two signatures: the average energy per particle,
and the shape of the time-of-flight density distribution.\index{Fermi
degeneracy}

Using eq.~\ref{YYZ:eq:NandE} we may write the average energy per
particle as 
\begin{equation} \label{YYZ:eq:E-per-N}
\frac{E}{N} = 3 k_B T \frac{f_4(\mathcal{Z})}{f_3(\mathcal{Z})}.
\end{equation}
The finite zero-temperature limit of eq.~\ref{YYZ:eq:E-per-N} is $3 E_F
/ 5$, corresponding to Fermi pressure \cite{YYZ:Pathria:SM}. By
comparison, the corresponding expression for the Boltzmann gas is $E/N
= 3 k_B T$, which tends toward zero at zero temperature.\index{Fermi
pressure}

The second signature of Fermi degeneracy is evident when the
observed fermion time-of-flight distribution is compared to the
predictions of the Boltzmann and Fermi-Dirac models.  The
latter is obtained by integration of eq.~\ref{YYZ:eq:positionSHO}
along the imaging line of sight.\footnote{When integrating Fermi
functions\index{Fermi function} over Gaussian degrees of freedom, it
is useful to note that $\int_{-\infty}^\infty dx \, f_n(C e^{-x^2}) =
\sqrt{\pi} f_{n+1/2}(C).$} \index{Fermi degeneracy}

Taking $z$ as the imaging direction, we find
\begin{equation} \label{YYZ:eq:Fermi-density-tof}
\tilde{n}(x, y, t) = \frac{N}{2 \pi \, r_x(t) \, r_y(t) \, 
f_3(\mathcal{Z})}
f_2\left(\mathcal{Z}\, \exp{\bigg[-\frac{x^2}{2 r_x^2(t)} -
\frac{y^2}{2 r_y^2(t)}}\bigg]\right),
\end{equation} 
where $r_i^2(t) = (\omega_i^{-2} + t^2)/\beta M$ is the cloud size in
the $i \in \{x,y,z\}$ direction after a time $t$ of free expansion. By
comparison, the spatial distribution for an expanding cloud of
classical particles is 
\begin{equation} \label{YYZ:eq:Boltzmann-density-tof}
\tilde{n}_\mathrm{cl}(x,y,t) = \frac{N}{2 \pi \, r_x(t) \, r_y(t)} 
\exp{\bigg[-\frac{x^2}{2 r_x^2(t)} -\frac{y^2}{2 r_y^2(t)}}\bigg], 
\end{equation} 
using the same definitions for $r_i(t)$.

\section{The atom chip}
\label{YYZ:sec:Orsay-chip}
The experiments described in this chapter were carried out at the
University of Toronto with an atom chip designed and fabricated in the
atom chip group at the Laboratoire Charles Fabry de l'Institut
d'Optique \cite{YYZ:Esteve:phd-thesis,YYZ:Aussibal:phd-thesis}. This
section details the conductor layout, material composition, and
supporting electrical and mechanical infrastructure for the chip.

\subsection{Chip construction and wire pattern}
\label{YYZ:sec:chip-pattern}
The atom chip consists of gold wires electroplated onto a cleaved
16~mm $\times$ 28~mm $\times$600~\um silicon substrate.  The chip was
patterned using photolithography, with the metal wires deposited by
evaporation and electroplating.  A 20~nm titanium adhesion layer and
200~nm gold seed layer were evaporated onto the SiO$_2$ surface oxide,
followed by solution electroplating of gold to a final wire height of
\mbox{6 \textmu m}. The final result is a mostly bare silicon
substrate hosting eight gold electrical contact pads connecting five
separate conductors on its surface \cite{YYZ:Esteve:phd-thesis}.  The
wires can sustain an operating DC current of 5 A, though we never
exceed 2~A in the central Z-wire.
\index{electroplating}\index{photolithography}

\begin{figure}[h]
\begin{center}
\begin{tabular}{cc}
\raisebox{3cm}{ 
 $\begin{array}{c}
 \includegraphics[angle=0,width=4cm]{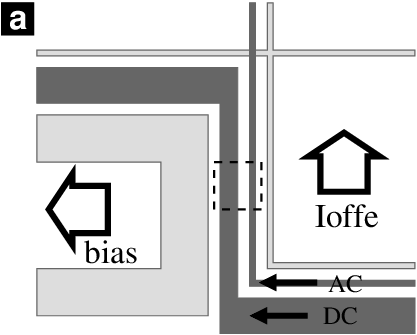}\\ \\
 \includegraphics[angle=0,width=6cm]{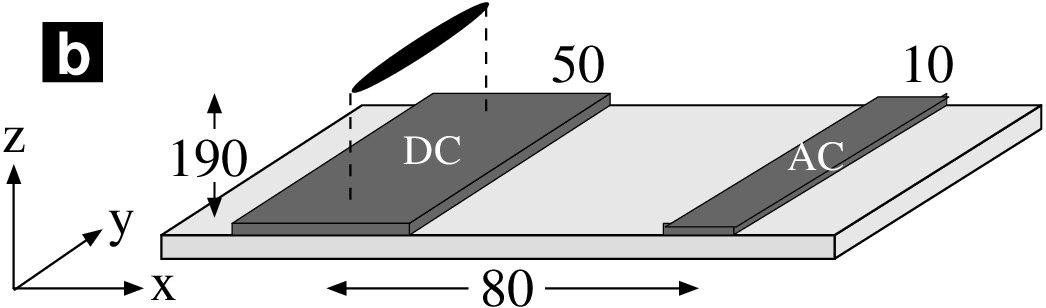} 
 \end{array}$ } 
&
\includegraphics[angle=0,width=4.5cm]{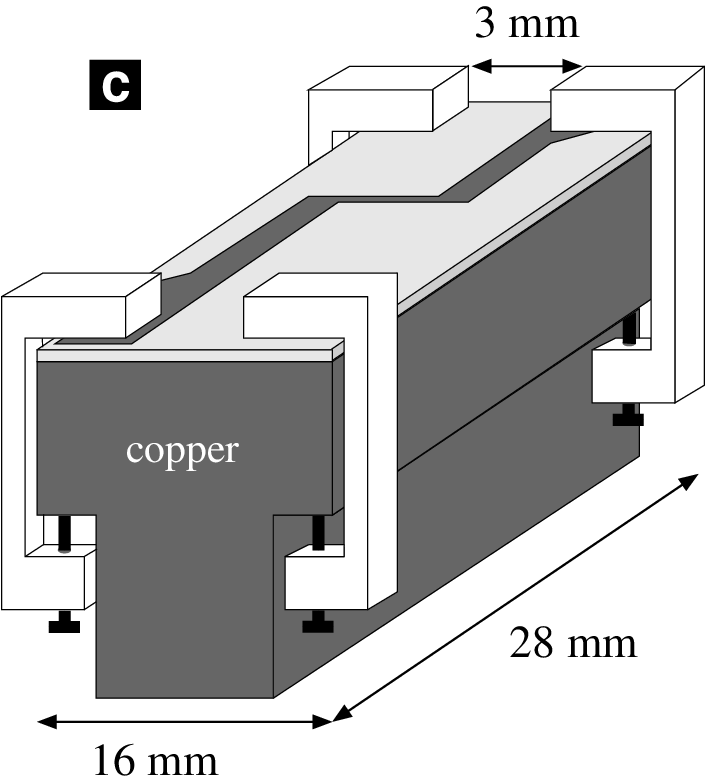}
\end{tabular}
\Caption{Atom chip conductor pattern and mounting schematic diagram.
\textbf{a}: Central atom chip conductor pattern, highlighting the Z-wire
(``DC'') and adjacent, thinner antenna wire (``AC'') in dark grey.  Thick
arrows indicate the directions of external, uniform magnetic fields.
\textbf{b}: Close-up view of dashed region in \textbf{a}, showing wire
widths, centre-to-centre separation, and location of the trapped
atomic cloud (black ellipsoid). All dimensions are in micrometres.
\textbf{c}: The atom chip is pressed onto a fly-cut copper block using
ceramic MACOR C-clamps (white).  Electrical connections are not shown.} 
\label{YYZ:fig:Orsay-chip-setup}
\end{center} 
\end{figure}

We use two of the five chip wires in the work presented here,
highlighted in dark grey in fig.~\ref{YYZ:fig:Orsay-chip-setup}a. The
central Z-wire forms the basis of our static micromagnetic trap. An
adjacent, thinner wire is used as a near-field antenna for delivering
RF and microwave fields to the trapped atoms.

\subsection{Electrical and mechanical connections} 
The chip is fastened to a copper support ``stack'' without glue or
screws, but by mechanical pressure only.  Custom-made MACOR ceramic
C-clamps press flattened strips of beryllium-copper foil onto the gold
contact pads for DC and RF electrical connections, while at the same
time pressing the chip onto the stack (see
fig.~\ref{YYZ:fig:Orsay-chip-setup}c).  The beryllium-copper foil
strips are in turn connected to ceramic-insulated copper wires which
deliver current from the air side into the UHV chamber and onto the chip.
The entire assembly is mounted vertically in the vacuum chamber with
the atom chip at the bottom, face down \cite{YYZ:Aubin:JLTP,
YYZ:Extavour:msc-thesis}.  The stack also acts as a heat-sink for the
atom chip, and was machined from oxygen-free high-conductivity (OFHC)
copper. \index{MACOR} \index{oxygen-free high-conductivity (OFHC)
copper}

\subsection{The Z-wire magnetic trap}
\label{YYZ:sec:Ztrap}
We use an anisotropic, Ioffe-Pritchard-type Z-wire microtrap
\cite{YYZ:Reichel:chipBEC-2002} throughout. The external DC bias
fields used for magnetic confinement are supplied by three pairs of
magnetic field coils mounted outside of the vacuum chamber.
With this setup we have achieved atom-surface distances between 80~\um
and 300 \textmu m, harmonic oscillation frequencies $\omega_{x,z}\sim
2\pi\times$200~Hz to $2\pi\times$2.5~kHz, $\omega_y \sim 2\pi
\times$32 Hz to 50 Hz, and trap depths as large as $k_B
\times$1.4~mK. \index{Ioffe-Pritchard trap}\index{trap depth}

The effects of wire surface roughness on the magnetic potential have
been studied extensively for atom chips of this type 
\cite{YYZ:Esteve:phd-thesis,YYZ:Esteve:frag,YYZ:Schumm:roughness}.
Though sub-micron-scale wire rugosity is not a problem at our 190~\um
working distance, we do observe three large and unexpected minima in
the longitudinal magnetic potential.  Though these ``defects'' prevent
us from reaching the largest trap anisotropies available in Z-traps,
we use them to our advantage during evaporative cooling. Magnetic
gradients applied along $y$ tilt the longitudinal potential so that
atoms preferentially fill the largest and deepest of these local
potential minima.  In this local trap, the longitudinal oscillation
frequency is two to three times larger than it would be in the absence
of the defect.  The increased mean oscillation frequency
$\bar{\omega}$ (see sec.~\ref{YYZ:sec:DFG-thermo}) and accompanying
gains in collision rate  allow us to evaporate to quantum degeneracy
more rapidly and efficiently than would otherwise be possible.

\section{Loading the microtrap} \label{YYZ:sec:loading}
Our experimental approach to creating a $\pf$ DFG on an atom chip is
motivated in large part by the scarcity of $\pf$, whose natural
isotopic abundance is only 0.012\%.  Even when using a potassium
dispenser with $\pf$   enriched to 5\%, we find that large
MOT beams are essential to capture enough $\pf$ in a background UHV
pressure compatible with magnetic trapping. Rather than using a
reflected surface MOT \cite{YYZ:Reichel:surface}, which would require
a mirror-coated chip of length $\sim$6~cm to accommodate our
4-cm-diameter MOT beams, we first load a large conventional MOT
several centimetres beneath the atom chip, then magnetically trap and
transport the atoms to the chip.  Once loaded into the Z-trap, we
sympathetically cool $\pf$ to quantum degeneracy with a $\re$
reservoir to minimize $\pf$ atom number loss.  This approach also
obviates the need for a mirror-coated atom chip.
\index{sympathetic evaporative cooling} \index{$\pf$}

In this section and the next, we review our laser cooling, magnetic
trapping and transport, chip loading, and evaporative cooling steps,
noting that further detail is available in
\cite{YYZ:Aubin:JLTP,YYZ:Aubin:chip-DFG, YYZ:Extavour:dual-degen}.  We
also add to our previous work with a discussion of the roles of trap
depth and trap volume in microtrap experiments, with particular
emphasis on chip loading in our setup.\index{trap volume}

\subsection{Laser cooling and magnetic transport to the chip}
\label{YYZ:sec:lasercool}
Both $\pf$ and $\re$ are initially trapped and cooled in a
dual-species MOT formed by six counterpropagating
4-cm-diameter beams centred 5~cm below the atom chip
\cite{YYZ:Aubin:JLTP, YYZ:Extavour:dual-degen}.  We use a
single-chamber vacuum system, and load the MOT directly from atomic
vapour created using a combination of dispensers and light-induced
atom desorption (LIAD).  During each MOT loading cycle, commercial
high-power 405 nm LEDs irradiate our
75~mm~$\times$~75~mm~$\times$~16~cm Pyrex vacuum cell for several
seconds to generate the atomic vapour. We use commercial Rb and
home-made K dispensers to replenish the $\pf$ and $\re$ coatings on
the interior walls of the UHV chamber as needed -- typically every few
days or weeks \cite{YYZ:Extavour:dual-degen}. By this method we
achieve $\pf$ and $\re$ atom numbers of roughly \mbox{$6\times10^6$}
and \mbox{$6\times10^8$} in the MOT.  \index{dual-species MOT}
\index{light-induced atom desorption (LIAD)}
\index{$\pf$}\index{$\re$}

After optical molasses cooling of $\re$, both $\re$ and $\pf$ are
optically pumped into the stretched internal magnetic hyperfine
states, $\ket{F=2,m_F=2}$ and $\ket{F=9/2,m_F=9/2}$,
respectively.  With all lasers extinguished, the mixture is
magnetically trapped in a quadrupole magnetic trap and transported
vertically to the surface of the atom chip using external coils. From
there atoms are smoothly transferred from the quadrupole magnetic trap
into the Z-trap located 190~\um from the chip surface, which has
harmonic $\pf$ oscillation frequencies $\omega_{x,z} =
2\pi\times$~823$\pm$7~Hz and $\omega_y = 2\pi\times$~46$\pm$1~Hz, and
a trap depth of \mbox{$\sim k_B\times$1.05~mK}.
\index{trap depth}\index{$\pf$}\index{$\re$}

\subsection{Loading bosons and fermions onto the atom chip}
\label{YYZ:sec:loading2}
Along with the advantages of high compression and fast collision rate
characteristic of atom chip microtraps comes a disadvantage: small
trap volume.  The volume occupied by a trapped gas
depends on its temperature, unlike in the uniform ``box'' potential 
we are accustomed to from thermodynamics.  The trap volume is not
typically discussed when creating ``macroscopic'' magnetic traps with
large coils, since their trap depths can be orders of magnitude larger
than is required to confine laser cooled atoms.  For microtraps,
however, the trap volume may limit the number of atoms that can
``fit'' into the trap, in a way that we will quantify in the following
subsections.  Our discussion points a clear route to larger atom
number, when it would be desirable.\index{trap depth}\index{trap
volume}

An atom is trapped when its energy is less than the trap depth $U_{td}$.
A good model of a thermalized gas in a trap-depth-limited trap is a
truncated Boltzmann distribution \cite{YYZ:Luiten:evap}, where
truncation occurs at $\eta$ times the temperature, i.e. $U_{td} = \eta
k_B T$. For the collision rates typical of atom chip traps, free
evaporation resulting from the limited trap depth occurs when $\eta
\gtrsim 3$; by contrast, \emph{efficient} evaporation occurs when $\eta
\gtrsim 5$. \index{trap depth}

The laser cooling discussed in sec.~\ref{YYZ:sec:lasercool} allows
atoms to be delivered to the chip at temperatures less than $U_{td}$.
However, our loading efficiency is typically 10\% or less, while
phase-space density is roughly preserved: we load roughly $2\times10^7$
$^{87}$Rb atoms and $2\times10^5$ $^{40}$K into the Z-trap. We will
attempt to explain the factors limiting this efficiency in the
following subsections. 

The initial loaded number of fermions places an upper bound on the
number of ultra-cold fermions we can produce. Furthermore, though we
load many more bosons than fermions, bosons are continually lost
during the evaporative cooling process, as described in
sec.~\ref{YYZ:sec:symp-cool}. The limited amount of ``refrigerant''
eventually limits the number of fermions that can be cooled, or their
final temperature. It is therefore important to understand our loading
process and ways in which it can be improved.

\subsection{Effective trap volume}\index{trap volume}
\label{YYZ:sec:Veff}
Our discussion here concerns the number, temperature, and density of a
gas at the start of evaporative cooling, having already been loaded
into the microtrap.  We will assume an initial phase-space density 
$\rho_0$, which is typically $\lesssim 10^{-5}$ for laser cooled atoms. At
the densities typical of this point in the experimental cycle, the
density distribution of the gas is well approximated as that of an
ideal non-degenerate gas.

From eq.~\ref{YYZ:eq:positionFiniteT}, the density distribution of an
ideal Fermi gas in three dimensions is 
$ n(\mathbf{r}) = \Lambda_T^{-3} f_{3/2} (\mathcal{Z} e^{-\beta 
U(\mathbf{r}) } )$,
where $U(0) \equiv 0$. The corresponding expression for ideal bosons
is obtained by using the thermodynamic Bose-Einstein function
$g_{3/2}(\cdots)$ in place of $f_{3/2}(\cdots)$ in this expression. In
either case, at low fugacity 
\begin{equation}
n(\mathbf{r})  \stackrel{\mathcal{Z} \ll 1}{\longrightarrow} 
\Lambda_T^{-3} \mathcal{Z} e^{-\beta U(\mathbf{r}) }.
\end{equation}
Integrating both sides of the equation, we recover the total atom number
\begin{equation}
N = n_0 \int e^{-\beta U(\mathbf{r})} d \mathbf{r},
\end{equation}
where the volume of integration is defined by $U<U_{td}$, and $n_0 =
n(0)$.

In analogy with a uniform gas, we define the effective volume 
\begin{equation}
V_\mathrm{eff} \equiv  \int e^{-\beta U(\mathbf{r})} d \mathbf{r},
\end{equation}
such that $n_0 = N/V_\mathrm{eff}$ \cite{YYZ:Luiten:evap}.

In the limit of $\eta \gg 1$, we can integrate the full Boltzmann
distribution to find the trap volume in several typical cases. For a
three-dimensional simple harmonic oscillator potential,\index{trap depth}
\begin{equation}
V^\mathrm{SHO}_\mathrm{eff} = \left(\frac{2 \pi}{M \bar{\omega}^2} 
\right)^{3/2} (k_B T)^{3/2},
\end{equation}
where $\bar{\omega}$ is the geometric mean oscillation frequency. For
the three-dimensional quadrupole (linear) trap, 
\begin{equation}
V^\mathrm{QT}_\mathrm{eff} = 8 \pi \bar{F}^{-3} (k_B T)^3,
\end{equation}
where $U \equiv | \mathbf{F} \cdot \mathbf{r} |$ and $ \bar{F}$ is the
geometric mean gradient.  For a simple three-dimensional box of side
$L$, $V_\mathrm{eff} = V = L^3$. Finally, a hybrid two-dimensional
quadrupole and one-dimensional box model gives
\begin{equation}
V^\mathrm{2QB}_\mathrm{eff} = 2 \pi L \bar{F}^{-2} (k_B T)^2.
\end{equation}

In all of the above cases, the effective volume has a power-law dependence 
$V_\mathrm{eff} = C_\delta T^\delta$.

\subsection{A full tank of atoms: maximum trapped atom number}
\label{YYZ:sec:number}
Using the effective volume, we can now relate the trapped atom number
to the initial phase-space density $\rho_0$, which is equivalent to
the degeneracy parameter $n_0 \Lambda_T^3$ for $\mathcal{Z} \ll 1$:
\begin{equation} \label{YYZ:eq:Ntrapped}
N = n_0 \Lambda_T^{-3} V_\mathrm{eff}(T).
\end{equation}
Since $k_B T \leq U_{td} / \eta$ by definition, we can write out the
explicit temperature dependence in eq.~\ref{YYZ:eq:Ntrapped} to find
\begin{equation} \label{YYZ:eq:Ntrapped2}
N \leq \rho_0 \left(\frac{M}{2 \pi \hbar^2}\right)^{3/2} C_\delta 
(U_{td}/\eta)^{\delta + 3/2},
\end{equation}
for a trap with a $\delta$ power law effective volume.

This equation shows us why the loaded atom number is typically smaller
in microtraps than macrotraps. First, the trap depth $U_{td}$ is
typically smaller, which reduces atom number with a power law as fast
as $U_{td}^{9/2}$ for a three-dimensional quadrupole. Second, even for
comparable trap depths, the stronger trapping strength of a microtrap
reduces $C_\delta$: $C_3 \propto \bar{\omega}^{-3}$ in the case of a
three-dimensional harmonic oscillator, for instance.\index{trap depth}

Eq.~\ref{YYZ:eq:Ntrapped2} also points to the importance of large
currents in trapping wires. Consider the case in which a
three-dimensional harmonic microtrap is formed above a single long
wire, such that $\omega_\perp \propto B_{0 \perp} / I$, where $B_{0
\perp}$ is the perpendicular bias field and $I$ is the wire current.
Since the trap depth increases linearly with $B_{0 \perp}$,
eq.~\ref{YYZ:eq:Ntrapped2} suggests that the maximum atom number at a
fixed  $\eta$ and phase-space density is $N_\mathrm{max} \propto I^2
B_{0 \perp} / \omega_z$. Assuming that the distance from the trapped
atoms to the chip surface is fixed, and that $\omega_z \propto
\sqrt{I}$, as is the case for our trap, $N_\mathrm{max} \propto
I^{5/2}$. Larger wire currents thus allow an increase in the number of
trapped atoms.\index{trap depth}

\subsection{Effect of geometry on loaded atom number}
\label{YYZ:sec:examples}
We now evaluate the trap volume and the expected maximum atom number
loaded into several well studied chip traps. We will start with the
earliest proposed traps, described in \cite{YYZ:WeinsteinLibbrecht} by
Libbrecht, and assume they are loaded with $^{87}$Rb in the
$|2,2\rangle$ state. For a single-loop quadrupole trap of radius
10~\um and using a 1~A current, the gradient is $5.4\times10^5$~G/cm
and the trap depth $k_B \times$21~mK.  Assuming the trap can be
loaded with $\eta \geq 4$ (corresponding to an initial temperature of
$\leq$~5~mK), the trap volume would be \mbox{$V_\mathrm{eff} \leq$ 310
\textmu m$^3$}. At an initial phase-space density of $10^{-6}$, $2
\times10^4$ atoms could be loaded into the trap.  However, a
quadrupole trap has a magnetic field zero at its centre, and is thus
unsuitable for trapping ultra-cold atoms.\index{trap depth}

The Ioffe ``(c)'' configuration of \cite{YYZ:WeinsteinLibbrecht}
consists of concentric half-loops with a 10~\um minimum diameter.
Using a 1~A current, the trap has a depth of \mbox{$k_B\times$1.3 mK}
and a curvature that gives $\bar{\omega} / 2 \pi \approx
94$~kHz. Although the trap is impressively strong, its effective
volume is only 0.4~\textmu m$^3$; less than one atom would be trapped at a
phase-space density of $10^{-6}$.  For this reason, larger trap
volumes than those of Libbrecht's pioneering geometries were required
to achieve quantum degeneracy in an atom chip microtrap.

Finally, let us consider the Reichel Z-trap
\cite{YYZ:Reichel:surface}. The potential at the centre of the trap is
harmonic, with a typical geometric mean frequency of
\mbox{$\bar{\omega} / 2 \pi \approx 300$ Hz} in our setup. The trap
depth is limited by the transverse applied field, for which a typical
value of 20~G gives \mbox{$U_{td} \approx k_B\times1.3$~mK}.
Assuming the trap is loaded at $\eta=4$, we find that the effective
volume is \mbox{$1.3\times10^7$~\textmu m$^3$}, and the maximum
trapped atom number \mbox{$1.2\times10^7$} at an initial phase-space
density of $10^{-6}$.  Although approximate, our calculation shows
that the Z-trap geometry is capable of loading six to seven orders of
magnitude more atoms than the Libbrecht geometry for the same initial
phase-space density. 

Furthermore, the calculation suggests that the loaded atom number in
our experiment is limited by trap depth and volume. For our geometric
mean field curvature of $3\times10^4$~G/cm$^2$ and initial
temperature of 300~\textmu K, the effective trap volume is $3\times
10^7$~\textmu m$^3$. One would expect $3\times10^7$ $^{87}$Rb atoms at
$\rho_0 \approx  10^{-6}$, and $3\times10^5$ $^{40}$K atoms at $\rho_0
\approx 4\times10^{-8}$. This is consistent with our observations,
even though the effective volume model is clearly simplified.
Nevertheless, the order-of-magnitude agreement demonstrates that we are
close to, if not at, the maximum possible number of loaded atoms,
given the phase-space density and temperature after magnetic transport
to the chip.  \index{trap depth}\index{trap volume}

\section{Rapid sympathetic cooling of a K-Rb mixture}
\label{YYZ:sec:symp-evap-dual-degen}
In this section we describe the sympathetic evaporative cooling of
$\pf$ with $\re$ in a microtrap, by which we produce a pure $\pf$
DFG or a dually-degenerate BEC-DFG mixture. Particular emphasis is
placed on the temperature dependence of the K-Rb scattering length,
and its effect on K-Rb rethermalization and evaporation efficiency. The
section concludes with a discussion of experimental signatures of
quantum degeneracy in DFGs. \index{sympathetic evaporative cooling}
\index{$\pf$}\index{$\re$} \index{evaporation efficiency}

\subsection{Forced sympathetic RF evaporation} \label{YYZ:sec:symp-cool}
We reach dual quantum degeneracy in $\pf$ and $\re$ via sympathetic RF
evaporative cooling of $\re$ in a Z-trap having \mbox{$B_0 \simeq$
2.6 G}, $\pf$ harmonic oscillation frequencies $\omega_{x,z} =
2\pi\times$~823$\pm$7~Hz and $\omega_y = 2\pi\times$~46$\pm$1~Hz, and
a trap depth of \mbox{$k_B\times$1.05~mK} \cite{YYZ:Aubin:chip-DFG,
YYZ:Roati:BEC-DFG, YYZ:Inguscio:collapse, YYZ:Jin:BECDFG,
YYZ:Sengstock:BEC-DFG}.  In our case, $\pf$ is cooled indirectly by
thermalizing elastic collisions with $\re$.  By sweeping the RF
evaporation frequency from 28.6 MHz to 3.65 MHz in as little as
\mbox{6 s}, we reach \mbox{$T/T_F \simeq 0.1 \,-\, 0.2$} with
\mbox{$\epsilon_F \simeq k_B \times 1.1$ \uK} and as many as \mbox{4
$\times$ 10$^4$} $\pf$ atoms, faster than has been possible in
conventional magnetic traps \cite{YYZ:Aubin:chip-DFG,
YYZ:Extavour:dual-degen}.\footnote{A $^6$Li DFG has been produced in
an all-optical setup in as little as 3.5~s \cite{YYZ:OHara:fastDFG}.}
This rapid evaporation to degeneracy is made possible by the
strong atom chip confinement.
\index{sympathetic evaporative cooling}\index{$\pf$}\index{$\re$}

\begin{figure}[h]
\begin{center}
\includegraphics[angle=0,width=8cm]{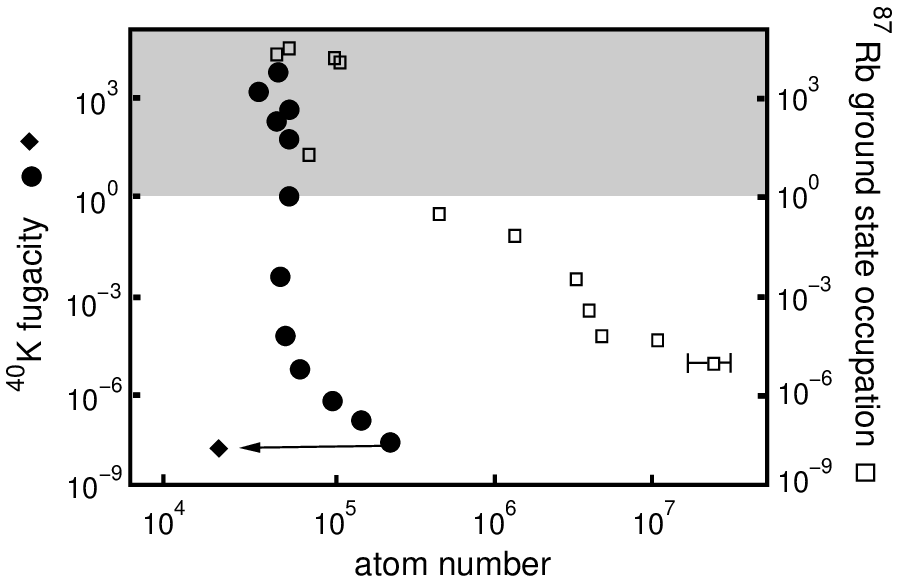}
\Caption{Sympathetic cooling to dual \mbox{$\pf$--$\re$} quantum
degeneracy.  Spin-polarized fermions without a bosonic bath cannot be
successfully evaporatively cooled (diamond).  However, if bosonic
$\re$ (squares) is evaporatively cooled, the fermionic $\pf$ is
sympathetically cooled (circles) by thermalizing elastic collisions
with $\re$.  The vertical axes indicate the evolution of
phase-space-density en route to dual quantum degeneracy ($\mathcal{Z}
\geq 1$) during evaporation: $\pf$ fugacity $\mathcal{Z}$ on the left,
and $\re$ ground state occupation on the right.  A typical run-to-run
spread in atom number is indicated on the right-most point; all
vertical error bars are smaller than the marker size. } 
\label{YYZ:fig:symp-evap} 
\end{center}
\end{figure}

As is evident in fig.~\ref{YYZ:fig:symp-evap}, the $\pf$ is cooled to
quantum degeneracy ($\mathcal{Z} \geq 1$) with only a
five-fold loss in atom number, while the $\re$ is evaporated with
log-slope efficiency \mbox{$-\partial [\log(\rho_0)]/\partial
[\log(N)] = 2.9 \pm 0.4$}, where $\rho_0$ is the peak phase-space
density.  When evaporating $\re$ alone to BEC, the evaporation
efficiency can be as high as \mbox{4.0 $\pm$ 0.1}; sacrificing some
evaporation efficiency, a more rapid evaporation can produce a BEC in
just 2 s. By contrast, for \mbox{$\pf$--$\re$} mixtures, we observe
that  RF sweep times faster than \mbox{6 s} are not successful in
achieving dual degeneracy.  The reason is that $\re$ and $\pf$
rethermalize more slowly than $\re$ alone, particularly during the
initial high-temperature stages of evaporation.  Direct measurements
of the $\pf$ and $\re$ temperatures indicate that $\pf$ thermalization
lags that of $\re$ despite an experimentally optimized RF frequency
sweep that is slower at higher temperatures and accelerates at lower
temperatures \cite{YYZ:Aubin:chip-DFG}.
\index{$\pf$}\index{$\re$}\index{fugacity}

\subsection{K--Rb cross-thermalization}
\label{YYZ:sec:KRb-cross-therm}
\begin{figure}[h]
\begin{center}
\includegraphics[angle=0,width=8cm]{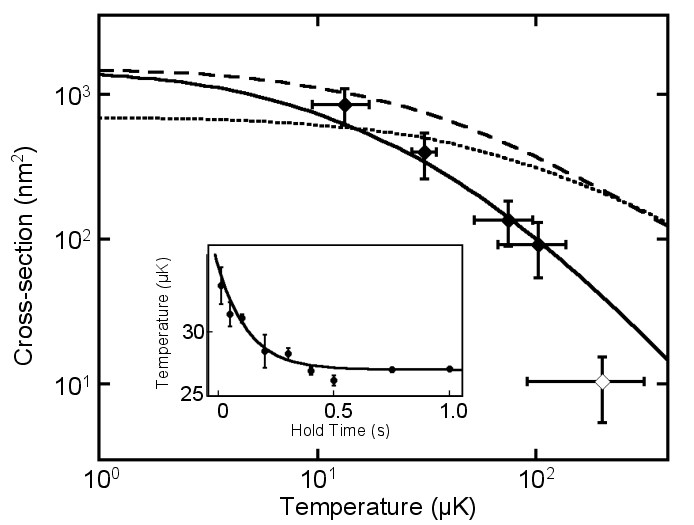}
\Caption{K--Rb cross-species thermalization.  Measurements of
$\sigma_{KRb}$ (diamonds) are compared with the s-wave-only (dashed)
and effective-range (solid) scattering models (see text).  For
reference, the s-wave $\sigma_{RbRb}$ is also shown (dotted).  Inset:
We measure the cross-thermalization by abruptly reducing the
temperature of $\re$ and watching the temperature of $\pf$ relax with
time.  The data shown has an asymptotic $\pf$ temperature of
27~\textmu K.  The highest temperature point (open diamond) did not
completely thermalize and lies off of the effective range prediction.
A more sophisticated analysis may be required for this point, owing to
severe trap anharmonicity at this high temperature.  The vertical
error bars are statistical (one standard deviation); the horizontal
error bars show the spread in initial and final $\pf$ temperature
during rethermalization.}

\label{YYZ:fig:sigma-K-Rb} 
\end{center}
\end{figure}

We have studied the \mbox{$\pf$--$\re$} inter-species scattering
cross-section $\sigma_{KRb}$ by measuring cross-thermalization rates at
temperatures between 10 \uK and 200~\uK \cite{YYZ:Aubin:chip-DFG}.  We
compare the results, shown in fig.~\ref{YYZ:fig:sigma-K-Rb}, to the
$\sigma_{KRb}$-vs.-temperature behaviour predicted by two scattering
models.  The simpler model, assuming only s-wave contact-interaction
scattering, predicts $\sigma_{KRb} = 4\pi a^2 / (1 + a^2 k^2) >
\sigma_{RbRb}$ throughout the stated temperature range, where $a =
a_{KRb}$ is the \mbox{$\pf$--$\re$} s-wave scattering length, and $k$
the relative wavevector in the centre-of-mass frame.  The second, more
detailed model is based on an effective-range atom-atom scattering
theory \cite{YYZ:Flambaum:scatt, YYZ:Anderlini:RbCs} and is
in good agreement with our measurements. 
\index{scattering cross-section} \index{$\pf$}\index{$\re$}

We attribute the observed reduction in scattering cross-section to the
onset of the Ramsauer-Townsend effect, in which the s-wave scattering
phase and cross-section approach zero for a particular value of
relative energies between particles \cite{YYZ:Townsend,
YYZ:Aubin:chip-DFG}.  Despite the high-temperature reduction in
cross-section, however, $\pf$ and $\re$ are relatively good
sympathetic cooling partners. By comparison, \mbox{$^6$Li--$\re$}
sympathetic cooling measurements \cite{YYZ:Silber:LiRb} suggest a
zero-temperature cross section approximately 100 times smaller than
$\sigma_{KRb}$; in other words, a maximum \mbox{$^6$Li--$\re$}
cross-section roughly equal to the lowest \mbox{$\pf$--$\re$} value we
measure. \index{Ramsauer-Townsend effect}
\index{scattering cross-section} \index{$\pf$}\index{$\re$}

\subsection{Density-dependent loss} \label{YYZ:sec:collapse}
The large and negative inter-species scattering length $a_{KRb}$
results in a strong \mbox{$\pf$--$\re$} attractive interaction.  At
low temperatures and high atomic densities, this interaction creates
an additional mean-field confinement that can lead to massive and
sudden losses during sympathetic cooling \cite{YYZ:Sengstock:BEC-DFG,
YYZ:Roati:BEC-DFG, YYZ:Inguscio:collapse}.  These studies of
density-dependent interaction-driven losses point to
boson-boson-fermion\footnote{Fermion-fermion-boson 3-body decay is
precluded by the Pauli exclusion principle.} 3-body decay as the
underlying mechanism for the collapse of the mixture.  This
effect ultimately limits the atom numbers in \mbox{$\pf$--$\re$}
mixtures.\index{$\pf$}\index{$\re$}

These effects manifest themselves in our experiment as $\pf$ and $\re$
number losses near the end of evaporation.  While we are able to
produce BECs of up to \mbox{$3 \times 10^5$} atoms when we work with
$\re$ alone, the simultaneous $\re$ and $\pf$ atom numbers are
restricted to at most \mbox{$\sim 10^5$} and \mbox{$4 \times
10^4$} respectively (see fig.~\ref{YYZ:fig:symp-evap}).
\index{$\pf$}\index{$\re$}\index{density-dependent 3-body loss}

\subsection{Required temperature} \label{YYZ:sec:temperature}
The reduction in the K--Rb elastic scattering cross-section discussed in
sec.~\ref{YYZ:sec:collapse} may lead the reader to wonder why we
started evaporative cooling at such a high temperature. The reason is
that the $\re$--$\re$ elastic collision rate required for efficient
evaporative cooling within our magnetic trap lifetime imposes a lower
bound on the initial temperature. In this section we will shift our
focus to \emph{single-species} collision rates in order to understand
this constraint.

Evaporative cooling requires a trap lifetime that is some multiple
(typically $10^3$) of $\gamma_\mathrm{coll}^{-1}$, where
$\gamma_\mathrm{coll} = n_0 \sigma v_\mathrm{r}$ is the collision rate
at the centre of the trap; $n_0$ is the central density, $\sigma$ is
the elastic scattering cross-section, and \mbox{$v_\mathrm{r} =
\sqrt{8 k_B T / \pi M}$} is the relative velocity of collision
partners \cite{YYZ:KetterleVanDruten:evap}. 

Since $\rho_0 \approx n_0 \Lambda_T^3$ at the start of evaporation
(see sec.~\ref{YYZ:sec:number}), we can express the central density in
terms of the phase-space density: 
\begin{equation}
\gamma_\mathrm{coll} = \frac{\sigma \rho_0 M}{\pi^2 \hbar^3} 
(k_B T)^2, 
\end{equation} 
which is independent of atom number. \index{scattering cross-section}

This relation can be used to give a lower bound on temperature in a
broad range of traps. Defining $\gamma_\mathrm{coll}^{min}$ as the
minimum scattering rate,
\begin{equation}
(k_B T)^2 \geq \gamma_\mathrm{coll}^{min} 
\frac{\pi^2 \hbar^3}{M \sigma \rho_0}.
\end{equation}
In the case of $^{87}$Rb, approximating $\sigma$ by its low
temperature limit $8 \pi a_s^2$, where $a_s$ is the s-wave scattering
length, we find that the minimum temperature at the beginning of
evaporative cooling is 
\begin{equation}
T_0^{min} = 300 \; \textrm{\textmu K} \times 
\left( \frac{10^{-6}}{\rho_0} \right)^{1/2} 
\left( \frac{\gamma_\mathrm{coll}^{min}}{150 \; \mathrm{s}^{-1}} 
\right)^{1/2}
\left( \frac{5.3 \; \mathrm{nm}}{a_s} \right),
\end{equation}
for \emph{any} loaded atom number or trap geometry. In our case, the
Rb--Rb elastic collision rate at the start of RF evaporation is
roughly 150 s$^{-1}$, the magnetic trap lifetime 5~s, and the
phase-space density $10^{-6}$ or slightly higher. Any adiabatic
cooling achieved by decompressing the trap would take us below this
minimum collision rate, and result in significant loss of evaporation
efficiency due to our relatively short trap lifetime.  We also observe
an increase in K--Rb 3-body loss (see sec.~\ref{YYZ:sec:collapse})
when evaporating with a higher collision rate in a more compressed
trap. For these reasons, we must start our evaporation in the regime where
the Ramsauer-Townsend effect is significant.
\index{evaporation efficiency} \index{Ramsauer-Townsend effect}

\subsection{Experimental signatures of Fermi degeneracy} 
\begin{figure} 
\begin{center}
\includegraphics[width=8cm]{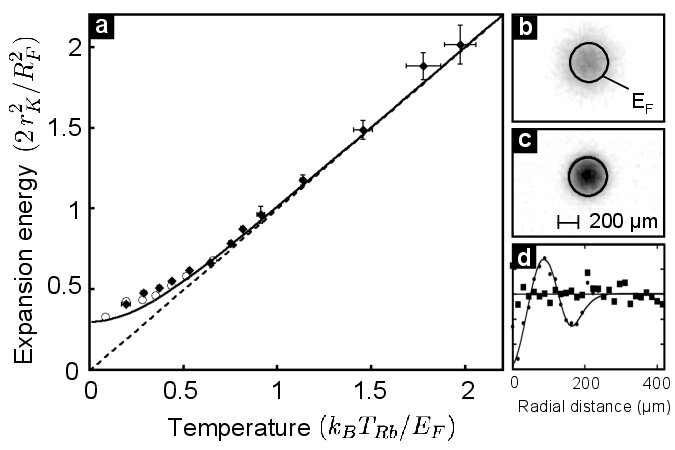}
\Caption{\textbf{a}: The apparent \emph{in situ} fermion temperature 
measured by gaussian fits to $\pf$ time-of-flight absorption data, 
plotted versus temperature of both thermal (diamonds) and
Bose-condensed (circles) $\re$. The data follows the expected
universal curve derived from ideal Fermi gas theory (solid line),
deviating from the corresponding Boltzmann prediction (dashed)
at temperatures below $T/T_F \approx 0.5$. \textbf{b,c}: Absorption
images for $T/T_F = 0.95$ and $T/T_F = 0.35$ respectively, overlayed
with black circles of radius $R_F$ rescaled after time-of-flight (see
sec.~\ref{YYZ:sec:dens-distn}). \textbf{d}: Residuals from radial
fits of gaussian (dots) and Fermi-Dirac (squares) envelope functions
to absorption data at $T/T_F =0.1$. The good Fermi-Dirac fit and poor
gaussian fit are evidence of a Fermi degenerate sample.}
\label{YYZ:fig:40K-expansion-energy}
\end{center} 
\end{figure}

Following the discussion in sec.~\ref{YYZ:sec:fermi-degen}, we assess
the degree of Fermi degeneracy in $\pf$ by fitting ideal Boltzmann and
Fermi gas theory to time-of-flight absorption data.  Unlike the
Boltzmann gas, whose spatial width tends toward zero as $T \to 0$,
Pauli exclusion \index{Pauli exclusion principle} results in a
finite-sized Fermi gas with a finite average momentum, even at $T=0$.
This Fermi pressure (see sec.~\ref{YYZ:sec:fermi-degen}) is evident in
fig.~\ref{YYZ:fig:40K-expansion-energy}a, in which the in-trap cloud
width deviates from the Boltzmann prediction for $T/T_F \lesssim 0.5$.
Absorption images taken at $T/T_F = 0.95$ and $T/T_F = 0.35$ are
overlayed with a circle indicating $E_F$ in
figs.~\ref{YYZ:fig:40K-expansion-energy}b and
\ref{YYZ:fig:40K-expansion-energy}c, demonstrating that the average
momentum of the Fermi gas plateaus at low temperature.
\index{Fermi degeneracy}\index{Fermi pressure}

In addition to the expansion energy, a second measure of the degree of
Fermi degeneracy is the deviation of the time-of-flight spatial
profile from the gaussian envelope predicted by Boltzmann statistics.
The gaussian fit is excellent at high temperatures, but fails at low
temperatures.  Fig.~\ref{YYZ:fig:40K-expansion-energy}d shows the
residuals of a gaussian fit using
eq.~\ref{YYZ:eq:Boltzmann-density-tof} to a gas at $T/T_F = 0.1$,
compared to those of a fit using equation
\ref{YYZ:eq:Fermi-density-tof}, which assumes Fermi-Dirac statistics.
Though the gaussian and Fermi-Dirac profiles are very similar,
careful analysis shows that the Fermi-Dirac function is a better fit,
with a reduced $\chi^2$ three times lower than the gaussian fit.
\index{Fermi degeneracy}

\section{Species-selective RF manipulation}\label{YYZ:sec:ssRF}
For atoms confined to a magnetic trap, an applied field oscillating at
radio frequencies (RF) can resonantly couple
adjacent $m_F$ states. The atomic system in the combined static and
time-varying magnetic fields can be described by a Hamiltonian with
new, uncoupled eigenstates -- the so-called ``adiabatic'' or ``RF-dressed''
states \cite{YYZ:ZobayGarraway:RF}.  The spatial dependence of these
states, and thus the spatial character of the trapping potential, can
be manipulated by varying the RF amplitude $B_{RF}$ and frequency
$\omega_{RF}$. In this section we discuss two useful regimes for
manipulating ultra-cold boson-fermion mixtures on atom chips using
these effects.  \index{RF-dressed potentials}

The first type of RF manipulation is species-selective evaporative
cooling.  As implied in sec.~\ref{YYZ:sec:symp-evap-dual-degen}, RF
manipulation can drive $\re$ spin-flip transitions to untrapped spin
states at the edge of the $\re$ dressed potential without inducing
loss in $\pf$, as a consequence of the explicit $m_F$ and $g_F$
dependence of the magnetic trapping potentials.
In the second type of RF manipulation, $B_{RF}$ is large enough that
Landau-Zener tunnelling between dressed states is suppressed and the
usual adiabatic condition is satisfied
\cite{YYZ:KetterleVanDruten:evap}.  Atoms adiabatically follow the
RF-dressed magnetic eigenstates and remain
trapped in a double-well potential \cite{YYZ:Colombe:phd-thesis,
YYZ:Lesanovsky:nonRWA1, YYZ:Lesanovsky:nonRWA2}. This effect was first
demonstrated on both thermal \cite{YYZ:Colombe-RFdressed,
YYZ:ZobayGarraway:RF} and quantum degenerate Bose gases
\cite{YYZ:Schumm:doubleBEC,YYZ:White:RF-dress} and is now a
well-established method for dynamically ``splitting'' an ultra-cold
Bose gas \cite{YYZ:Schumm:doubleBEC,YYZ:Hofferberth-RF1,
YYZ:Lesanovsky:RF, YYZ:GBJo:dw1, YYZ:vanEs:RF-dress}.  We have added
to this work by demonstrating the simultaneous creation of a $\re$
double well and a $\pf$ single well \cite{YYZ:Extavour:dual-degen}.
This result demonstrates that the adiabatic potentials experienced by
each species can be dramatically different in an applied RF field,
owing to the different values of $g_F$ for $\re$ and $\pf$.
\index{RF-dressed potentials} \index{adiabatic potentials}
\index{species-selective}\index{Landau-Zener transition} \index{$\pf$}

\subsection{Sympathetic RF evaporation}
\label{YYZ:sec:weak-RF-dress}
In this section we consider a simple sympathetic cooling scenario,
typical in our experiments; an RF ``knife'' is used to evaporatively
cool $\re$, which then acts as a refrigerant for the
$\pf$.\footnote{Microwave magnetic fields at 6.8~GHz may also be used
to species-selectively evaporate $\re$ via hyperfine transitions
\cite{YYZ:Silber:LiRb}.} Ideally, this scheme would causes no loss of
$\pf$ atoms; here we explore to what extent this is
true.\index{species-selective}\index{$\pf$}

A critical parameter in evaporative cooling is $\eta = U_{td}/k_B T$
(see sec.~\ref{YYZ:sec:loading2}).  The larger the choice of $\eta$,
the fewer atoms are removed from the trap, since a smaller fraction of
the kinetic energy distribution lies above $U_{td}$
\cite{YYZ:KetterleVanDruten:evap}.  In sympathetic cooling of $\pf$ by
$\re$, we would ideally like the RF field to have no effect on the
$\pf$ trap depth.  However, we shall see that for our scheme the RF
knife \emph{does} limit the trap depth for $\pf$, though the effect is
small. This effect may be quantified by evaluating $\eta$ for $\pf$ in
our cooling scheme.\index{$\pf$}

We begin by considering atoms in spin states $\ket{F,m_F}$ in a static
(``DC'') magnetic trap, whose potential is $U(\mathbf{r}) = m_F g_F
\mu_B B_{DC}(\mathbf{r})$.  The minimum magnetic field amplitude is $B_0
\equiv B_{DC}(0)$ using the notation $B_{DC}(\mathbf{r}) \equiv
|\mathbf{B}_{DC}(\mathbf{r})|$. We apply a weak, sinusoidally
time-varying magnetic field of amplitude $B_{RF}$ and frequency
$\omega_{RF}$.  If $\hbar \omega_{RF}>g_F \mu_B B_0$, then atoms in
state $m_F$ at positions $\mathbf{r}_1$ for which
\begin{equation} \label{YYZ:eq:B1} 
g_F \mu_B
B_{DC}(\mathbf{r}_1)=\hbar\omega_{RF} 
\end{equation} 
can undergo spin flips to adjacent $m_F$ states, including untrapped
ones. This imposes an effective trap depth $U_{td}=g_F m_F
\mu_B(B_{DC}(\mathbf{r}_1) - B_0)$ on these atoms, since any atom
with energy above $U_{td}$ is ejected from the trap.  
Therefore, we can write
\begin{equation} \label{YYZ:eq:etakBT} 
U_{td} \equiv \eta k_B T = m_F(\hbar\omega_{RF}-g_F \mu_B B_0) 
\end{equation} 
for a gas of atoms in state $\ket{F,m_F}$ in thermal equilibrium at
temperature $T$.

For two species in thermal equilibrium in the same
magnetic trap, we can write down simultaneous equations like
\ref{YYZ:eq:etakBT} for both species. From these, we infer the
relationship
\begin{equation} \label{YYZ:eq:etas}
\eta_{K} = \left( \frac{m_{F}^{(K)}}{m_{F}^{(Rb)}} \right) \eta_{Rb}
+ m_{F}^{(K)} \left[g_{F}^{(Rb)} - g_{F}^{(K)}\right] 
\frac{\mu_B B_0}{k_B T}.
\end{equation}
between the two $\eta$ parameters in the case of $\pf$
and $\re$, using the fact that $\omega_{RF}$ and $B_0$ are common to
both species. This convenient form highlights the role of $m_F$ and
$g_F$ in the RF evaporation of a mixture of atomic species.
\index{$\pf$}

Moving now to a more specific case, we note that typical sympathetic
cooling ramps are done with $\re$ in the $\ket{2,2}$ state, and the
$\pf$ in $\ket{9/2,9/2}$, since this mixture is
stable with respect to many inelastic collisions
\cite{YYZ:Roati:BEC-DFG}. Eq.~\ref{YYZ:eq:etas} yields
\begin{equation} \label{YYZ:eq:etas-KRb} 
\eta_K =  \frac{9}{4}\eta_{Rb} +
\frac{5}{4} \frac{\mu_B B_0}{k_B T},   
\end{equation}
from which we immediately conclude that
$\eta_K> 9\eta_{Rb}/4$ for all values of temperature.  Having
$\eta_K > \eta_{Rb}$ ensures that $\re$ can be evaporated
without inducing significant loss in the $\pf$ population, as required
for efficient sympathetic evaporative cooling.  

\begin{figure}[h]
\begin{center}
\includegraphics[width=8cm]{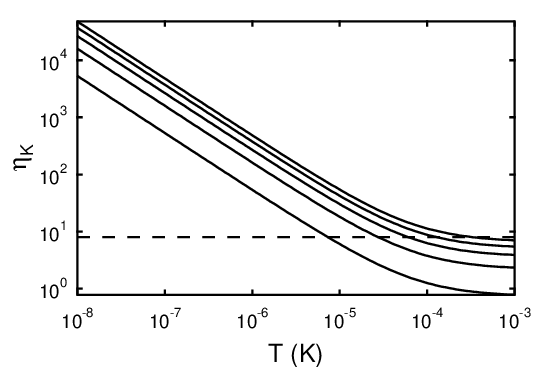}
\Caption{Evaporation parameter $\eta_K$ for $\pf$ atoms in a typical
sympathetic cooling ramp with $B_0 = 5.7 G$ and 
$\eta_{Rb} \approx 8$ (dashed line, see text).  The curves represent
different magnetically trappable values of $m_F^{(K)}$ ($1/2$ to
$9/2$, from bottom to top).  $\eta_K$ rises sharply as evaporation
proceeds and the temperature decreases, while $\eta_{Rb}$ remains
constant.}
\label{YYZ:fig:eta-vs-T} 
\end{center} 
\end{figure}

For a typical experimental value of $\eta_{Rb} \approx 8$, which
is roughly constant throughout the evaporation, fig.~\ref{YYZ:fig:eta-vs-T}
demonstrates that $\eta_K$ rises sharply for all trappable $m_F^{(K)}$
states as sympathetic evaporation proceeds and the $\pf$ temperature
decreases. $\pf$ spin-flip losses are most likely to occur when
$\eta_K$ is smallest.  In our experiments, this occurs at the
beginning of evaporation, when $B_0=$ 5.7~G, $T \sim $ 300 \uK
\cite{YYZ:Aubin:chip-DFG, YYZ:Extavour:dual-degen}, and $\eta_K
=$~8.1.

We can also use eq.~\ref{YYZ:eq:etas-KRb} to compare $\eta_K$ and
$\eta_{Rb}$ in the scenario in which all the $\re$ is evaporated away
to leave a pure $\pf$ DFG.  Taking $\eta_{Rb} \to 0$ and working at a
typical pure-DFG temperature $T =$~220~nK \cite{YYZ:Aubin:chip-DFG},
we find $\eta_K >$~220 for all trappable $\pf$ sublevels.  This
confirms that $\re$ can be fully and selectively ejected from the trap
without causing any significant $\pf$ loss.

Finally, we note that it is possible in principle to evaporate $\pf$
without inducing $\re$ loss. For RF frequencies $\hbar\omega_{RF} \leq
g_F^{(Rb)}\mu_B B_0 $, only $\pf$ atoms are ejected from the trap, at
positions $\mathbf{r}_2$ such that $\hbar\omega_{RF} = g_F^{(K)} \mu_B
B(\mathbf{r}_2)$. Following the derivation of eq.~\ref{YYZ:eq:etakBT}
, we can express the $\pf$ trap depth in this scenario as
$U_{td}^{(K)} = m_F^{(K)} [g_F^{(Rb)}
- g_F^{(K)} ] \mu_B B_0$. In our experiment, using $B_0 =5.7$~G and
  the stretched states of $\pf$ and $\re$, $U_{td} = 5 \mu_B B_0 / 4
\simeq k_B\times480$~\textmu K.  Thus, we expect to be able to
evaporate $T \lesssim 480$~\uK $\pf$ clouds without affecting
the $\re$ population.

\subsection{Species-selective double wells}
\label{YYZ:sec:strong-RF-dress}
In this section we discuss the species-selective nature of RF-dressed
double wells.  In a two-species mixture, this effect permits the
simultaneous formation of a double-well potential for $\re$ and
single-well potential for $\pf$ \cite{YYZ:Extavour:dual-degen}, and
vice versa.  We use an RF amplitude $B_{RF} \ll B_0$, for which
the effective adiabatic potentials\footnote{For $B_{RF} \approx B_0$,
eq.~\ref{YYZ:eq:Ueff-RWA}, which relies on the rotating-wave
approximation (RWA), is no longer valid \cite{YYZ:Lesanovsky:nonRWA1,
YYZ:Lesanovsky:nonRWA2, YYZ:Hofferberth:nonRWA}.} may be written
\cite{YYZ:Schumm:doubleBEC, YYZ:Hofferberth-RF1, YYZ:Lesanovsky:RF}
\index{RF-dressed potentials}\index{species-selective} 
\index{adiabatic potentials}
\begin{equation} \label{YYZ:eq:Ueff-RWA} 
U_{\mathrm{eff}}(\mathbf{r}) = m_F' \sqrt{ \delta(\mathbf{r})^2 + 
\Omega(\mathbf{r})^2 } 
\end{equation} where 
\begin{align} \label{YYZ:eq:delta-Omega} 
\delta(\mathbf{r})&=\mu_B g_F B_{DC}(\mathbf{r})-
\hbar \omega_{RF} \\
\Omega(\mathbf{r})&=\mu_B g_F B_{RF\perp}(\mathbf{r}) / 2,
\end{align}
$B_{DC}(\mathbf{r})\equiv|\mathbf{B}_{DC}(\mathbf{r})|$ is the
static magnetic field amplitude, $B_{RF\perp}(\mathbf{r})$ is the
amplitude of the $B_{RF}$ component which is perpendicular to
$\mathbf{B}_{DC}(\mathbf{r})$ at point $\mathbf{r}$, and $m_F'$ is the
new, effective magnetic quantum number.  The parameters
$\delta(\mathbf{r})$ and $\Omega(\mathbf{r})$ can be identified as the
local detuning and Rabi frequency of the RF field, respectively.
Using eq.~\ref{YYZ:eq:Ueff-RWA}, we can calculate effective adiabatic
potentials for $\re$ and $\pf$ and use them to illustrate the
simultaneous creation of single- and double-well potentials in a
\mbox{$\pf$--$\re$} mixture.\index{adiabatic potentials}\index{double
well}

\begin{figure}[h]
\begin{center}
$\begin{array}{cc}
\includegraphics[width=4.5cm]{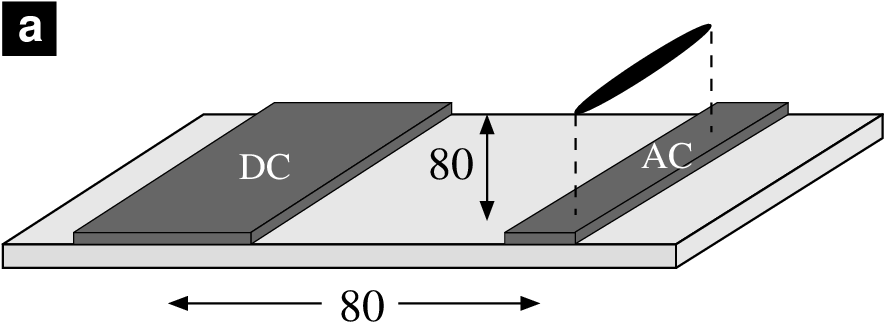}
\includegraphics[width=5.5cm]{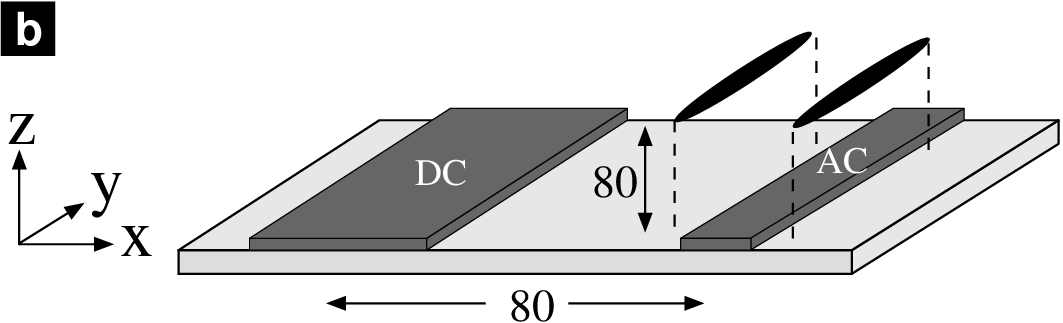}
\end{array}$
\Caption{Atoms confined to single-well (\textbf{a}) and double-well
(\textbf{b}) potentials induced by RF manipulation 80~\textmu m above
the chip surface.}
\label{YYZ:fig:split-geometry}
\end{center} 
\end{figure}

For clarity, we consider the formation of the $\re$ double well and $\pf$
single well separately.  Our starting point is a
$\ket{9/2,9/2}$--$\ket{2,2}$ \mbox{$\pf$--$\re$} mixture confined to a
static, anisotropic harmonic Z-trap directly above the RF wire (see
fig.~\ref{YYZ:fig:split-geometry}a) with $B_0 =$ 1.214 G,
$\omega_{x,z} = 2\pi\times$ 1.23~kHz and $\omega_y = 2\pi\times$ 13.7
Hz.\index{double well}

\begin{figure}[h]
\begin{center}
$\begin{array}{cc}
\includegraphics[angle=0,width=5.2cm]{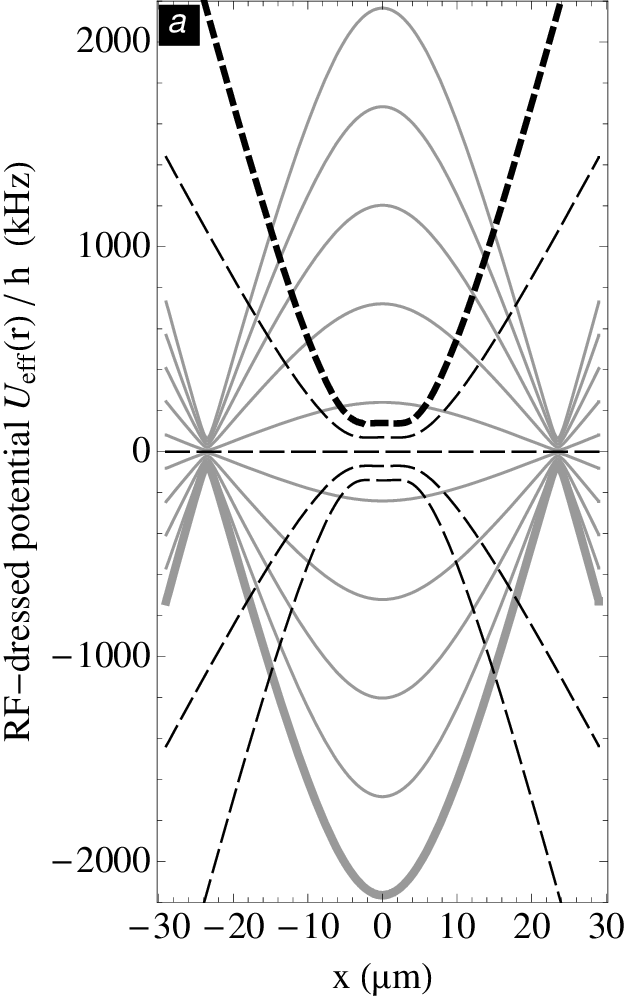} &
\raisebox{4.3cm}{
  \begin{tabular}[t]{c}
  \includegraphics[angle=0,width=5cm]{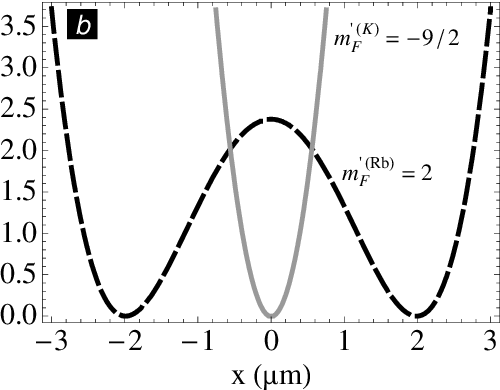} \\
  \includegraphics[angle=0,width=5cm]{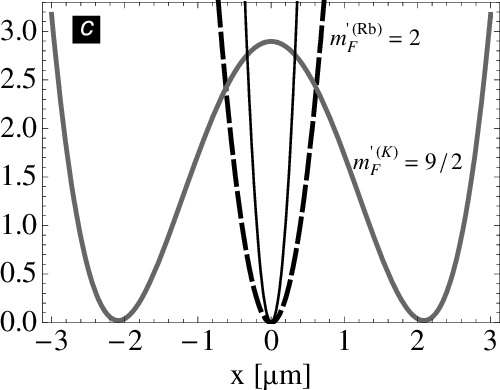}
  \end{tabular} }
\end{array}$
\Caption{Simultaneous adiabatic dressed potentials for $\re$ and
$\pf$. \textbf{a}: The $\re$-double-well case. $\re$ (black) and $\pf$
(grey) effective dressed potentials are plotted as a function of the
spatial coordinate $x$ (see fig.~\ref{YYZ:fig:split-geometry}).
Each curve corresponds to a single value of $m_F'$ for $\re$ and
$\pf$, with the upper-most curves corresponding to \mbox{$m_F'=2$} and
\mbox{$m_F'=9/2$} respectively.  $\re$ atoms populate their upper-most
\mbox{$m_F'=2$} dressed state (thick, dashed), while $\pf$ atoms
populate their lower-most \mbox{$m_F'=-9/2$} single well dressed
state (thick, grey). \textbf{b}: A closer view of the $\re$ double
well and $\pf$ single well, plotted together on a single vertical
$U_{\mathrm{eff}}/h$ axis in units of kHz. Both curves have been shifted
vertically to align their potential minima at zero kHz.  \textbf{c}: A
closer view of the $\pf$ double well and $\re$ single well, with
vertical axis similar to \textbf{b}.  The dressed $\re$ single well
(black, dashed) deviates slightly from the undressed single well
(solid, thin black), illustrating the slight loss of radial trap
curvature.} 
\label{YYZ:fig:dressed-potentials}
\end{center} 
\end{figure}

\paragraph{Rb double well}
An RF field with initial frequency $\omega_{RF}=2\pi\times 800$~kHz
and detuning $\delta^{(Rb)} = -50$~kHz is applied by
ramping up its amplitude from zero to the final value $B_{RF}=200$~mG.
A potential barrier is formed at $\mathbf{r} = 0$ by sweeping the RF
frequency through the resonant point $\delta^{(Rb)}(0) = 0$. As the RF
field is applied, each undressed state is adiabatically connected to
one dressed state; here $m_F^{(Rb)} = 2$ is connected to
$m_F'^{(Rb)}=2$, shown as the upper-most black curve in
fig.~\ref{YYZ:fig:dressed-potentials}a. After sweeping to a final RF
frequency \mbox{$\omega_{RF} = 2\pi\times 860$~kHz}, the barrier height is
$h\times$2.4~kHz and the $x$-direction double well separation is
4~\um (see fig.~\ref{YYZ:fig:dressed-potentials}b).  The $m_F'^{(Rb)}$
level repulsion at the double well minima is 70~kHz, sufficient to
prevent Landau-Zener spin flips at our working temperatures \mbox{$T
\lesssim$ 1 \textmu K}.  The $\re$ population thus remains trapped in
the $m_F'^{(Rb)} =2$ dressed level.  
\index{Landau-Zener transition}\index{double well}

\paragraph{K single well} The trapping potential for the $\pf$ atoms
is affected in a very different way for the same magnetic field
configuration.  In our current example
(fig.~\ref{YYZ:fig:dressed-potentials}a,b), the detuning for $\pf$ is
positive; $\delta^{(K)}(0)=2\pi\times$482~kHz at the trap minimum.
Near $\mathbf{r}=0$, the RF dressing adiabatically connects
$m_F^{(K)}=9/2$ to $m_F'^{(K)}=-9/2$.  Since $[\delta^{(K)}(0)]^2 \gg
[\Omega^{(K)}(0)]^2$, the potential curvature near $\mathbf{r}=0$ is
largely unaffected by the RF coupling (see equation
\ref{YYZ:eq:Ueff-RWA}).  More quantitatively, the dressed states are
most deformed where $\delta^{(K)} (\mathbf{r}) \approx 0$, which
corresponds to \mbox{$x\simeq\pm 23$ \textmu m}, and a potential
energy of roughly 110~\uK above the local minimum at $\mathbf{r}=0$.
Since our experiments are typically conducted with \mbox{$T\lesssim$ 1
\textmu K}, we can be satisfied that the $\pf$ potential retains its
original form near $\mathbf{r} = 0$ without inducing any $\pf$ loss.

One important feature of this single- and double well arrangement is
that the $\re$ double well separation and barrier height may be tuned
over a wide range by adjusting $\omega_{RF}$ and $B_{RF}$ without
affecting the shape of the $\pf$ potential.\index{double well}

An obvious extension of the work described here and in
\cite{YYZ:Extavour:dual-degen} would be to reverse the roles of boson
and fermion, creating a double well for fermions overlapped with a
single well for bosons.  The magnetic hyperfine structure of $\pf$ and
$\re$ makes this possible in a \mbox{$\pf$--$\re$} mixture, but with
slightly different results than in the $\re$-double-well case.
\index{double well}

Following the $\re$-double-well example of the preceding section, here
we sweep the RF frequency from $\omega_{RF} = 2\pi\times338$ kHz to
$\omega_{RF} = 2\pi\times383$ kHz. In the same static trap with $B_0
= 1.214$ G, $\delta^{(K)}$ changes sign from $-50$~kHz to $+5$~kHz,
while $\delta^{(Rb)}$ remains negative throughout. This creates a
$\pf$ double well in the $m_F'^{(K)} =9/2$ state with $x$-direction
well separation \mbox{$\sim 4$~\textmu m}, barrier height
\mbox{$h\times$2.9 kHz} at $\mathbf{r}=0$, and $m_F'^{(K)}$ level
repulsion $\sim$~70~kHz at \mbox{$x = \pm 2.1$~\textmu m}.  In
contrast to the $\re$-double-well scenario, here \emph{both} $\pf$ and
$\re$ adiabatically follow their respective upper-most dressed levels,
which exhibit their strongest spatial deformation near the trap
centre.  While $\pf$ experiences a double well potential, the $\re$
$m_F'^{(Rb)}=2$ potential is a single well with slightly reduced
radial curvature from the initial, undressed $m_F=2$ potential, as
shown in fig.~\ref{YYZ:fig:dressed-potentials}c.  \index{double well}

In addition to the species-selectivity of this process, it should be
emphasized that atom chips are particularly well-suited to creating
adiabatic dressed state potentials due to the proximity of the atoms
to chip wire RF antennae. The double-wells described in this section
were created using RF Rabi frequencies \mbox{$\Omega \sim$ 100 - 200
kHz}, though we can achieve values as large as 1~MHz with tens of
milliamperes rms in the chip wire antenna. By comparison, achieving
$\Omega \approx$~1~MHz with an air-side RF antenna would require a
circular coil of radius 3~cm and 3 turns bearing 10~A rms of AC
current. \index{double well}

\section{Fermions in an optical dipole trap near an atom chip}
\label{YYZ:sec:fermions-ODT}
In our discussion of fermions on atom chips thus far, we have focused
on spin-polarized $\pf$ in magnetic traps.  In this section we
describe an extension of our atom chip capabilities with the
incorporation of an external crossed-beam optical dipole trap skimming
the surface of the atom chip.  Optical trapping enables the use of 
any and all hyperfine and Zeeman spin states in our experiments, as
well as more flexible control of the magnetic field environment. These
added features allow us to work with strongly interacting $\pf$ spin
mixtures, while retaining the atom chip benefits of near-field RF and
microwave manipulation and rapid evaporation to Fermi degeneracy. 
\index{Fermi degeneracy}\index{optical dipole trap}

\subsection{Optical trap setup}
\begin{figure}[h]
\begin{center}
\begin{tabular}{cc}
\includegraphics[angle=0,width=3.5cm]{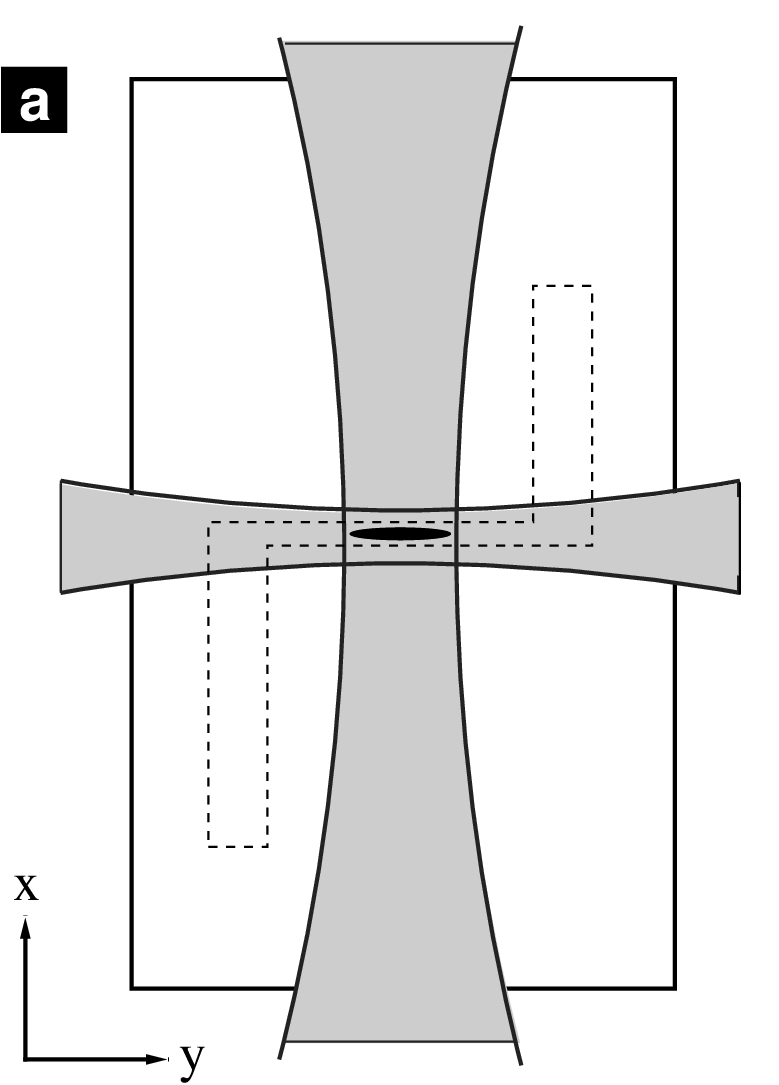} &
  \raisebox{2.5cm}{ 
  $\begin{array}{c}
   \includegraphics[angle=0,width=4cm]{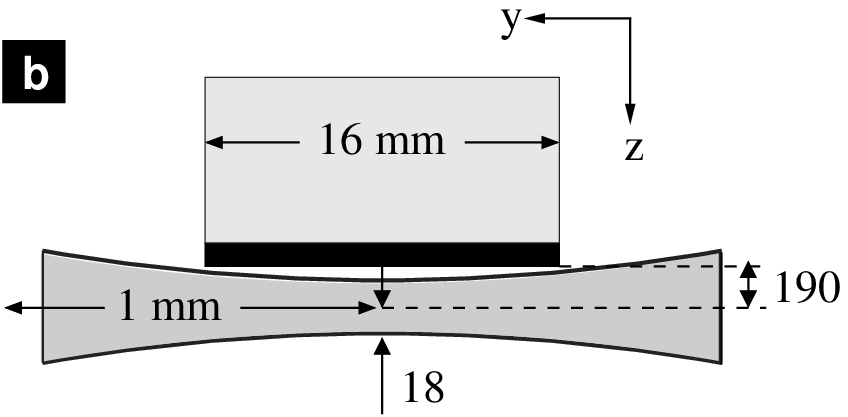} \\ \\ 
   \includegraphics[angle=0,width=4cm]{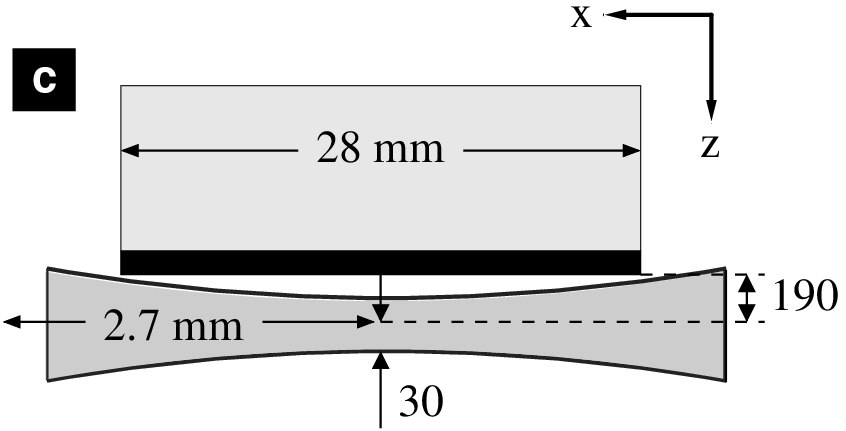}  
  \end{array}$ }
\end{tabular}
\Caption{Alignment of the dipole beams to the atom chip.
\textbf{a}: Bottom view, showing the alignment with the Z-wire and the
trapped atoms (black ellipse). \textbf{b,c}: Side-views of beams
skimming 190~\textmu m below the atom chip surface.  Beam
waists and Rayleigh ranges are indicated, along with the outer
dimensions of the atom chip (black) and copper mounting block (light
grey).  Dimensions are in units of micrometres unless otherwise
indicated.}\index{Rayleigh range}
\label{YYZ:fig:chip-ODT}
\end{center} 
\end{figure}

Two optical trapping beams are generated using the output of a single
500~mW \mbox{Nd:YAG} laser operating at \mbox{$\lambda =$ 1064 nm},
and directed along the $x$ and $y$ directions through the vacuum cell
and beneath the atom chip using air-side optics (see
fig.~\ref{YYZ:fig:chip-ODT}). Ideally, the foci would be near enough
to the chip surface to allow sufficient RF and microwave coupling, but
not so near that the optical potential is degraded by light scattering
off the edges of the chip as the beams are focussed.  The beams should
also be aligned to the existing Z-trap position for efficient transfer
from the Z-trap into the dipole trap. To satisfy these constraints we
focus the two beams at roughly 190~\um beneath the chip surface, with
$1/e^2$ waists $w_0 \sim$ 18~\um and \mbox{30 \textmu m}, and Rayleigh
ranges \mbox{$z_0 = \pi w_0^2 / \lambda \sim$ 1.0~mm} and 2.7~mm,
respectively, as shown in figs.~\ref{YYZ:fig:chip-ODT}b and
\ref{YYZ:fig:chip-ODT}c.  \index{Rayleigh range}

\subsection{Loading the optical trap}
We load the optical trap by ramping off the Z-trap magnetic fields and
ramping on the optical beams one at a time. First, the $y$-direction
peak beam intensity is increased linearly from zero to \mbox{$\sim 2.9
\times 10^7$ mW/cm$^2$} in 100~ms.  The Z-trap is then ramped off in
50~ms.  Finally, the $x$-direction peak beam intensity is linearly
ramped from zero to \mbox{$\sim 8.5 \times 10^6$ mW/cm$^2$} in 100~ms.  

Unlike when loading the Z-trap from the macroscopic magnetic trap (see
sec.~\ref{YYZ:sec:lasercool}), the mode matching between the Z-trap
and the initial single-beam optical trap is excellent.  This allows us
to load the optical trap with no observable loss in atom number, but
not without introducing some heating: \mbox{6 $\times$ 10$^3$} atoms
in the Z-trap at $T/T_F \simeq$ 0.56 are all transferred into the
optical trap, after which $T/T_F \simeq $1.

\subsection{Microwave and RF manipulation} 
\begin{figure}[h]
\begin{center}
$\begin{array}{ccc}
 \includegraphics[angle=0,width=4.5cm]{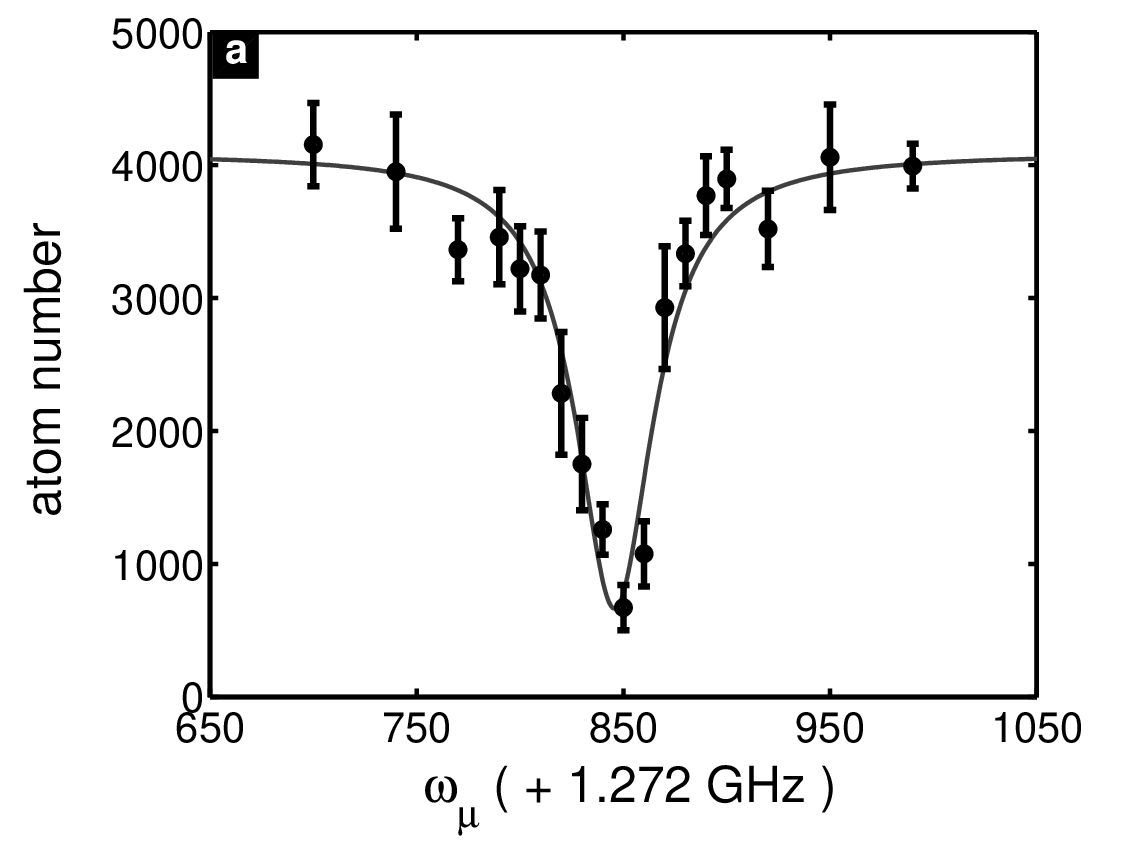} &
 \raisebox{5mm}{
  \includegraphics[height=2.7cm]{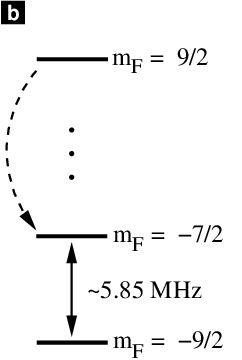} } & 
 \includegraphics[angle=0,width=4.5cm]{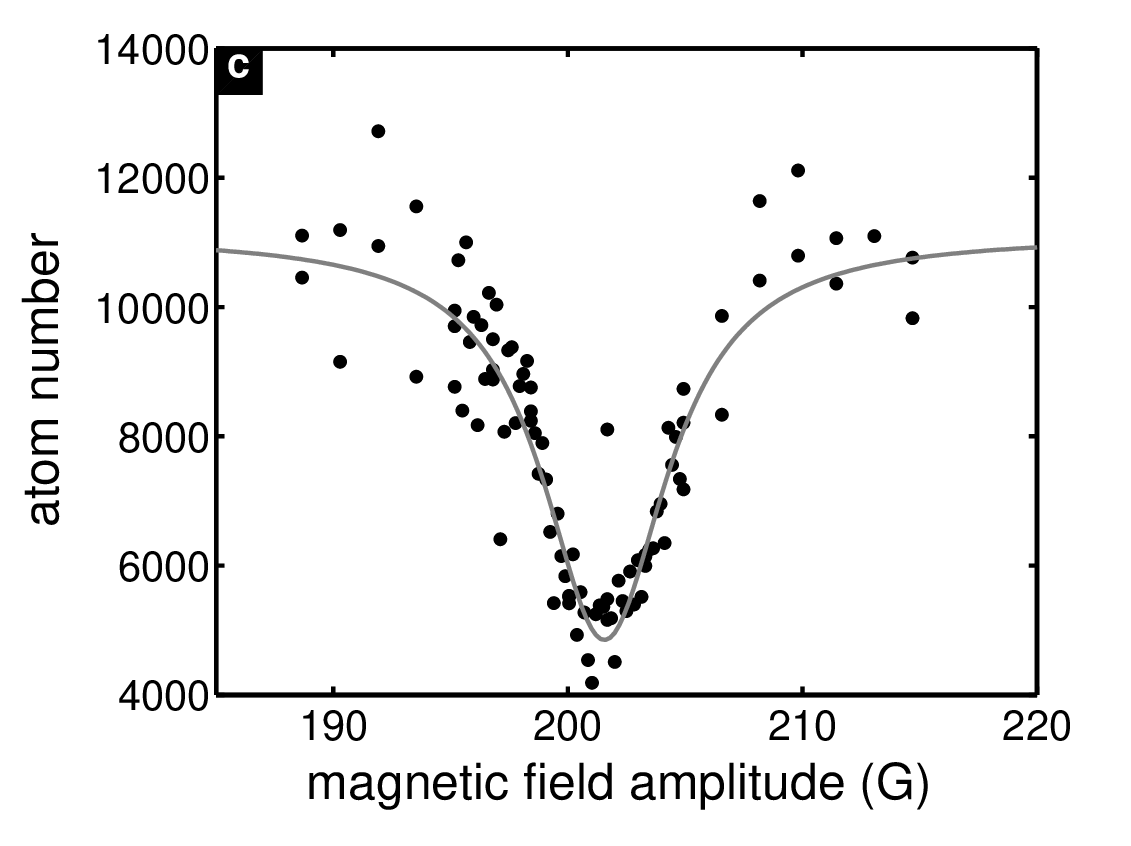} 
\end{array}$
\Caption{\textbf{a}: Microwave-induced atom number $\pf$ loss from a
magnetic Z-trap. \textbf{b}: Schematic Zeeman level diagram of
the $\pf$  $F=9/2$ hyperfine ground state, indicating the
microwave adiabatic rapid population transfer (dashed arrow) and RF
pulse (solid arrow) used to create a $\ket{9/2,-7/2}$ --
$\ket{9/2,-9/2}$ spin mixture (see text). \textbf{c}: 3-body loss of
spin-mixed $\pf$ from an optical trap is due to an enhancement of the
scattering cross-section induced by the Feshbach resonance near
202~G.} 
\label{YYZ:fig:microwave-manip} 
\end{center} \end{figure}

With the atoms now trapped solely by the optical trap, we can use the
antenna wire (see sec.~\ref{YYZ:sec:chip-pattern}) to apply microwave
radiation with amplitude $B_\mu$ and frequency \mbox{$\omega_\mu \sim
2\pi\times$1.3 GHz} to drive transitions between $\pf$ hyperfine
ground states.  To test and calibrate our microwave system, however,
we first observed the effects of microwave radiation on magnetically
trapped $\pf$. By scanning the microwave frequency near 1.3~GHz in a
Z-trap with magnetic minimum $B_0=$~5.22~G, we observe a loss feature
in the trapped atom number, corresponding to hyperfine transitions
between the initial trapped state $\ket{9/2,9/2}$ and the final,
untrapped $\ket{7/2,7/2}$ state (see
fig.~\ref{YYZ:fig:microwave-manip}a).

Microwave and RF manipulation also allow us to create an interacting
Fermi gas in the optical trap with a
$\ket{9/2,-9/2}$--$\ket{9/2,-7/2}$ spin mixture after loading a
spin-polarized sample from the Z-trap in state $\ket{9/2,9/2}$.  Atoms
are transferred from $\ket{9/2,9/2}$ to $\ket{9/2,-7/2}$ by rapid
adiabatic transfer.  In a 20~G external magnetic bias field, we sweep
the microwave frequency from below the $\ket{9/2,9/2} \to
\ket{7/2,7/2}$ transition (\mbox{$\omega_\mu = 2\pi\times$1235.8
MHz}) to above the $\ket{7/2,-7/2} \to \ket{9/2,-7/2}$ transition
(\mbox{$\omega_\mu = 2\pi\times$1336.2 MHz}) in 500~ms. This is
followed by a 5~ms RF frequency sweep from \mbox{$\omega_{RF}
=2\pi\times$5.80 MHz} to \mbox{$2\pi\times$5.90 MHz} (see
fig.~\ref{YYZ:fig:microwave-manip}b), creating an equal spin mixture
of $\ket{9/2,-9/2}$ and $\ket{9/2,-7/2}$.
\index{rapid adiabatic transfer}

The $\pf$--$\pf$ interaction strength may be dramatically increased
using the well-known Feshbach resonance in a DC magnetic field of
roughly 202 G \cite{YYZ:Loftus:40K-Feshbach}. Using external coils, we
apply a magnetic bias field and monitor the $\pf$ population as a
function of this field. We have observed the Feshbach resonance in our
atom chip setup as a strong loss feature near 202 G, as shown in
fig.~\ref{YYZ:fig:microwave-manip}b.  The atom number loss is
attributed to 3-body decay induced by the strong interactions near the
Feshbach resonance.

\section{Discussion and future outlook} \label{YYZ:sec:outlook}
The union of Fermi gases and atom chips is an important step forward
in degenerate quantum gas research. In this chapter, we have
described our work producing the first DFG on an atom chip.  Using a
large dual-species MOT, $\pf$ and $\re$ are captured and loaded onto
an atom chip in numbers large enough to enable sympathetic evaporative
cooling to DFG and dual DFG-BEC quantum degeneracy.  The strong
confinement and large inter-species collision rate afforded by the
micromagnetic trap permits an evaporation to quantum degeneracy in
as little as 6~s, faster than had been previously possible in magnetic
traps. We have also demonstrated species-selective double-well
potentials in a $\pf$--$\re$ mixture, as well as the creation of a
strongly-interacting $\pf$ DFG in a crossed-beam optical dipole trap
skimming the surface of the atom chip.\index{dual-species MOT}
\index{species-selective}\index{double well} 
\index{optical dipole trap}

We also note several disadvantages associated with using micromagnetic
traps for fermions. Magnetic traps are not ideal for working with the
spin-mixtures required for an interacting DFG, since spin-changing
collisions can populate magnetically untrapped spin states, causing
loss.  We address this limitation by incorporating a crossed-beam
optical dipole trap into our atom chip setup. The relatively small
trap volumes of microtraps also limits the number of target $\pf$ and
refrigerant $\re$ atoms that can be loaded into the chip trap.  Even
at our maximum achievable trap depth, these limits on the number of
refrigerant atoms ultimately limit the DFG to \mbox{4$\times10^4$}
atoms at \mbox{$T/T_F \gtrsim 0.1$} in our system. By comparison,
larger, colder, interacting DFGs are routinely produced in
spin-independent optical potentials \cite{YYZ:manyBody-RMP:2008}.
\index{optical dipole trap}\index{trap depth}\index{trap volume}

Despite these drawbacks, atom chips remain a useful and versatile tool
for ultra-cold Fermi gas research. We close this chapter with
descriptions of some of our ongoing and future research directions
with a new atom chip: double-well potentials; interacting DFGs and
DFG-BEC mixtures; and optical probes and traps.\index{double well}

\paragraph{New atom chip}
We have recently adopted a richer atom chip design, which improves on
its predecessor (described in sec.~\ref{YYZ:sec:Orsay-chip}) in three
important areas. First, the new conductor layout allows a wider
selection of micromagnetic traps with greater tunability of the
oscillation frequencies.  We are also now able to generate RF
double-well potentials at arbitrary angles and distances from the chip
surface, as in \cite{YYZ:Hofferberth-RF1}.
Second, new near-field antenna wire geometries and high-frequency
electrical feed-throughs improve microwave and RF impedance matching,
reducing coupling losses between the air-side and the chip.  Finally,
all electrical connections to the chip are made on the back side of
the substrate, increasing the optical access for probe and trapping
beams compared to its predecessor. Designed and fabricated at the
University of Toronto \cite{YYZ:Jervis:msc-thesis}, the chip consists
primarily of silver conductors that were patterned by photolithography
and evaporatively deposited onto an aluminum nitride
substrate.\index{aluminum nitride}

\paragraph{Bose-Fermi mixtures and 
double-well potentials}
The species-selectivity of RF-dressed double-well potentials may be
useful in studying boson-fermion interactions in ultra-cold atomic
mixtures.  The strong attractive interaction between $\pf$ and $\re$,
known to impede sympathetic cooling in \mbox{$\pf$--$\re$} mixtures,
depends on the inter-species collision rate and related peak $\pf$ and
$\re$ number densities \cite{YYZ:Roati:BEC-DFG, YYZ:Inguscio:collapse,
YYZ:Sengstock:BEC-DFG, YYZ:Aubin:chip-DFG}.  Adiabatic RF manipulation
could be used to reduce the $\re$ peak density by decompressing the
$m_F'^{(Rb)}=2$ effective potential at the centre of the trap during
sympathetic RF evaporation.  Ideally, the RF-dressed
\mbox{$\pf$--$\re$} collision rate would be small enough to avoid a
density-driven collapse, but still large enough to maintain good
inter-species rethermalization for sympathetic
cooling.\index{RF-dressed potentials}

These potentials are also amenable to the study of phase coherence in
the \mbox{$\pf$--$\re$} mixture as the potential barrier is raised.
Recent studies of phase-coherent RF splitting of $\re$ BECs on atom
chips \cite{YYZ:Hofferberth-RF1, YYZ:GBJo:dw1, YYZ:Schumm:doubleBEC}
and in optical traps \cite{YYZ:Gati:BEC-BJJ-2006,
YYZ:Esteve:BEC-squeezing} have focused on interactions and tunnelling
in BEC-in-double-well systems.  An interesting extension of this work
would be to assess the effects of a background, unsplit fermion
``bath'' on the tunnelling dynamics and coherence properties of this
system.  

\paragraph{Light scattering by a DFG} 
An inhibition of optical scattering is predicted to occur in DFGs when
$E_F \gtrsim E_R$, where $E_R$ is the recoil energy
\cite{YYZ:DeMarco:light-scatt, YYZ:Busch:DFG-light-scatt}.  The
optical scattering rate depends on the availability of atomic recoil
states (``final states''), which is constrained due to Pauli blocking
in the filled or nearly-filled Fermi sea \index{Fermi sea}of a trapped
DFG.  The large oscillation frequencies and Fermi energies available
in anisotropic, harmonic atom chip microtraps ($E_F \propto
\bar{\omega}$, see sec.~\ref{YYZ:sec:DFG-thermo}) offer a promising
route toward measurements of this effect (see
\cite{YYZ:Shuve-JHT:DFG-light-scatt} and references therein).
\index{light scattering}

\setlength{\bibindent}{4mm} 

\bibliographystyle{atchip}
\bibliography{refs-Toronto-fermichip-Wiley} 


\end{document}